\documentclass[journal]{IEEEtran}

\usepackage[T1]{fontenc}
\usepackage{amsmath}
\usepackage{color,soul}
\usepackage{algorithm,algpseudocode}
\usepackage{graphicx}
\graphicspath{ {./Figures/} }
\usepackage[cmintegrals]{newtxmath}
\usepackage{float}
\usepackage{cuted}

\allowdisplaybreaks

\begin{document}
	
	\title{Energy-Efficient Backscatter Aided Uplink NOMA Roadside Sensor Communications under Channel Estimation Errors}
	\author{Asim~Ihsan, Wen~Chen,~\IEEEmembership{Senior Member,~IEEE,} Wali~Ullah~Khan, Qingqing Wu,~\IEEEmembership{Senior Member,~IEEE,} and Kunlun Wang 
	\thanks{(Corresponding author: Wen Chen.)}
	\thanks{Asim Ihsan and Wen Chen are with Department of Information and Communication Engineering, Shanghai Jiao Tong University, Shanghai 200240, China.	Email: \{ihsanasim;wenchen\}@sjtu.edu.cn.}
	\thanks {Wali Ullah Khan is with the Interdisciplinary Centre for Security, Reliability and Trust (SnT), University of Luxembourg, 1855 Luxembourg City, Luxembourg (Emails: waliullah.khan@uni.lu).}
    \thanks {Qingqing  Wu  is  with  the  State  Key  Laboratory  of  Internet  of Things  for  Smart  City,  University  of  Macau,  Macau,  China  (e-mail: qingqingwu@um.edu.mo).}
    \thanks {Kunlun Wang is with the School of Communication and Electronic Engineering, East China Normal University, Shanghai 200241, China (e-mail: klwang@cee.ecnu.edu.cn).}}
		\maketitle
\begin{abstract}
This work presents non-orthogonal multiple access (NOMA) enabled energy-efficient alternating optimization framework for backscatter aided wireless powered uplink sensors communications for beyond 5G intelligent transportation system (ITS). Specifically, the transmit power of carrier emitter (CE) and reflection coefficients of backscatter aided roadside sensors are optimized with channel uncertainties for the maximization of the energy efficiency (EE) of the network. The formulated problem is tackled by the proposed two-stage alternating optimization algorithm named AOBWS (alternating optimization for backscatter aided wireless powered sensors). In the first stage, AOBWS employs an iterative algorithm to obtain optimal CE transmit power through simplified closed-form computed through Cardano’s formulae. In the second stage, AOBWS uses a non-iterative algorithm that provides a closed-form expression for the computation of optimal reflection coefficient for roadside sensors under their quality of service (QoS) and a circuit power constraint. The global optimal exhaustive search (ES) algorithm is used as a benchmark. Simulation results demonstrate that the AOBWS algorithm can achieve near-optimal performance with very low complexity, which makes it suitable for practical implementations.
\end{abstract}


\IEEEpeerreviewmaketitle

\section{Introduction}
\label{sec:introduction}
\IEEEPARstart{A}{n} effective traffic monitoring system is essential for ITS to surveil the prevailing conditions across the road. ITS demands a wide range of sensors on the road to manage transportation in an intelligent, seamless, safe, and secure manner. These sensors could monitor traffic patterns, determine optimum traffic routing, identify traffic accidents, and perform environmental measurements \cite{M. Habibzadeh}. In such a large-scale network, it is very essential to collect and transmit information with little power consumption. These sensors consist of sensing, computation, and communication components that consume power. Backscatter communication is an emerging technology that provides a single infrastructure for jointly sensing and transmitting data with microwatt levels of power consumption \cite{U. S. Toro}. 

Backscatter communications provide batteryless connectivity through RF energy harvesting \cite{A. Bletsas}. This technique is very different than the general energy harvesting techniques where devices harvest energy for the operation of active RF transmissions \cite{Q. Wu, Q. Wu 2}. On the contrary, backscatter communication is low complexity and low power technique that does not require any active RF transmission component. It allows transmitters to transfer their information to the backscatter reader (BR) installed on receivers by reflecting and modulating the RF carrier signals of CE. Such backscattering at transmitters is done through mismatching the impedance at the input of the antenna which results in varying reflection coefficients \cite{X. Lu}. Its transmission consumes very little energy as compared to the conventional radio and its operation without active RF component results in much simpler and uncomplicated circuits. The implementation of the backscatter technique for the wirelessly powered sensors has been limited because of its limited coverage. Backscatter communication based on bistatic architecture has been proposed to mitigate this limitation \cite{A. Bletsas,P. N. Alevizos,J. Kimionis}. In bistatic backscatter communication, the CE is dislocated from BR, which results in a more flexible network configuration and reduces the near-far effect. Therefore, it is much more suitable for future ITS to use bistatic backscatter aided roadside sensors.

\subsection{Related Literature}
Recently, backscatter communication has shown great potential for large-scale massive internet of things (IoT) networks \cite{A. E. Mostafa}. To achieve low latency, high spectral, and energy efficiency, NOMA is the key technology to serve a large number of roadside sensors. The integration of NOMA with backscatter communication has proven great potential for collecting information from multiple sensors through the same sub-channel in a non-orthogonal manner \cite{S.Zeb, F.Jamel}. The information of multiple roadside sensors can be multiplexed on the same sub-channel by tuning the value of the reflection coefficient of each roadside sensor to a different value. After tuning their reflection coefficients, multiple roadside sensors in the same cluster can be separated through the power domain. Then, the BR can decode the information of each roadside sensor by employing the power difference of their signals. The authors in \cite{S.Zeb}, used the combination of time-division multiplexing access (TDMA) with NOMA as hybrid access schemes for monostatic backscatter communication. They improved the network performance in terms of throughput and outage probability. In \cite{Q. Zhang}, authors investigated the ergodic rate and outage probability for backscatter NOMA system that integrates downlink NOMA communication with backscatter device. The performance of NOMA-enabled backscatter communication in terms of successful decoding of an average number of bits at the reader is analyzed in \cite{J. Guo}. In \cite{A. Farajzadeh}, authors explored NOMA backscatter communication for UAV application by optimizing UAV altitude and trajectory. Author in \cite{G. Yang}, jointly optimize backscatter time and power reflection coefficients for throughput maximization of NOMA bistatic communication network. EE of the symbiotic system that integrates downlink NOMA communication with backscatter device is maximized in \cite {Y. Xu}. 

 So far following research contributions explored backscatter-aided vehicular communications. The overview of different  use  cases  of NOMA-enabled backscatter aided  6G  vehicular  networks is presented in \cite{W and A}. In \cite{Wali}, authors formulate the joint optimization of BS and RSU power allocation for efficient backscatter enabled vehicular communication with NOMA. They considered a downlink scenario for cooperative NOMA communication, in which multiple RSUs are assisting BS to multicast the information to the vehicles along with backscatter tag to vehicles communication. A novel learning-based optimization framework for backscatter aided heterogeneous vehicular networks is presented in \cite{Wali 3}. The authors in \cite{Felisberto Pereira}, analyzed the integration of passive sensors and uplink backscatter communication for vehicular technology. They demonstrated the use of piezoelectric transducers for passive sensing along with uplink backscatter communication for pedestrians' safety and validated it through their experiments. The contribution in \cite{V. Hansini}, proposed backscatter-aided secure vehicle-to-vehicle communication for managing parking situations in VANETs. The backscatter technology for vehicular positioning is investigated in \cite{K. Han}. 
 
\subsection{Motivations and Contributions}
Roadside sensors collect large amounts of information and send it to central computing servers through roadside units (RSUs) or base stations (BSs) for analysis \cite{X. Liu sensors}. Besides, RSUs can also transfer sensed information to the vehicles that exist in their coverage area. For instance, sensors detect the pedestrians in the crosswalk and transmit it to RSUs through backscatter communication. Then, RSUs disseminate this information to surrounding vehicles in its coverage \cite{Felisberto Pereira}. Hence, such an interconnected transport system results in better decision-making that leads to the improved safety of the environment. \cite{U. S. Toro}. As the roadside sensors need to collect and transmit different useful information to the RSUs in an energy-efficient manner. Therefore, backscatter communications can be employed by the roadside sensors to send their sensed information to the nearest RSU in the uplink scenario \cite{W and A, Felisberto Pereira}. Besides, Obtaining CSI is crucial for the performance analysis of backscatter communications. In practice, it is very challenging for the passive transmitter to guarantee the accuracy of  CSI all the time \cite{Y. Zhang}. Furthermore, the realization of backscatter enabled uplink sensor communications will demand advanced energy-efficient resource allocation (RA) frameworks. Their RA is incredibly challenging due to the diverse quality-of-service (QoS) requirements and its strong underlying dynamics. In literature, we believe that there is no energy-efficient RA optimization framework for NOMA-enabled backscatter aided uplink roadside sensors communication with channel uncertainties. Therefore, in this article, we proposed a novel alternating optimization framework for EE maximization of wireless powered backscatter aided uplink sensors communications for future generation ITS under imperfect CSI. Our major contribution is summarized as follows,

\begin{table}
	\fontsize{9}{11}\selectfont
	\centering
	\caption{THE LIST OF DIFFERENT ABBREVIATIONS AND THEIR DEFINITIONS}
\begin{tabular}{l|l} 
		\hline 
		Acronym & Definition  \\
		\hline 
	    AOBWS & Alternating optimization for backscatter aided \\ & wireless powered sensors.\\
		BR & Backscatter reader.\\
		BS & Base station. \\
		CCFP & Concave-convex fractional programming.\\
		CE & Carrier emitter.\\
		CSI & Channel state information. \\
		EE & Energy efficiency. \\
		ES & Exhaustive search.\\
		IoT & Internet of things.\\
		ITS & Intelligent transportation system. \\ 
		KKT &  Karush–Kuhn–Tucker.\\
		NOMA & Non-orthogonal multiple access. \\
		OCETP & Optimal CE transmit power.\\
		OFDMA & Orthogonal frequency-division multiple access.\\
		QoS & Quality of service. \\
		RF & Radio frequency.\\
		RSU & Road side unit. \\
		SIC & Successive interference cancelation.\\
		TDMA & Time-division multiplexing access. \\ 
		UAV & Unmanned aerial vehicles. \\
		VANET & Vehicular ad-hoc network. \\
		\hline
	\end{tabular}
\end{table}

\begin{itemize}
	\item A novel energy-efficient alternating optimization framework for NOMA-enabled wireless powered uplink sensor communication under imperfect CSI is proposed for ITS. The EE maximization problem is formulated under various QoS requirements of bistatic backscatter communications for roadside sensors. The EE is maximized under channel uncertainties by optimizing transmit power of CE and reflection coefficient of roadside sensors. The formulated problem is solved into two stages, which yields the proposed AOBWS algorithm. 
	\item The proposed two-stage AOBWS algorithm provides optimal EE performance in very low computational complexity, which makes it suitable for practical implementations. The detailed complexity analysis of the proposed algorithm and benchmark algorithms is presented in the complexity analysis subsection.
	\item The efficacy of the proposed AOBWS algorithm is analyzed through numerical simulations and is compared with the global optimal ES algorithm as a benchmark. From obtained results, it is observed that AOBWS can achieve desired EE performance with affordable computational complexity for practical implementations.
		
\end{itemize}
The rest of the paper is organized as follows: The system model  is provided in Section II. In Section III, we discussed the problem formulation with its solution for the energy efficient wireless powered uplink sensor communication under channel uncertainties for ITS while Section IV presents the numerical results to verify the efficacy of the proposed AOBWS algorithm. Finally, Section V provides concluding remarks of our analysis and discuses the future work. The definition of different acronyms and symbols used in this
paper are respectively defined in Table I and II.

\section{System Model}
This work considers NOMA-enabled backscatter communication for wireless powered passive sensors in ITS. The system model consists of a CE, multiple backscatters aided roadside sensors, and BR installed on RSU as depicted in Figure 1. Roadside sensors are multiplexed in various clusters, where each cluster consists of $K$ sensors. Practically, It is desireable to have two or three sensor per cluster for low decoding complexity and to guarantee timing constraints \cite{A. W. Nazar}. Roadside sensors harvest RF energy from the RF carrier signal emitted by CE \cite{F. Jameel} and backscatter their information to the BR. It works in two operational modes, namely the transmission mode ($T_t$) and energy harvesting mode ($T_h$). These two modes of roadside sensor constitute one slot such that $T_t+ T_h =1$ \footnote{Optimizing time coefficients of transmission mode $T_t$ and energy harvesting mode $T_h$ of roadside sensor can further enhance the EE performance, however, it is left for our future work to focus here on optimizing CE transmit power and reflection coefficient of roadside sensors under channel uncertainties.}, where $T_t >0$ and $T_h>0$. In transmission mode, each cluster of roadside sensors backscatters the RF carrier signal of CE to transmit its sensed information to the BR. This backscattering is carried out with the help of an RF transistor. It reflect the incident carrier RF signal with altered phase and magnitude through mismatching the impedence at the input of the antenna which results in varying reflection coefficients. In the energy harvesting mode, roadside sensor harvests energy from incident RF signal instead of reflecting it for transferring information. This harvested energy is then reserved in the battery and is utilized to power its circuitry. For a more detailed description of backscatter enabled sensors, please refer to \cite{A. W. Nazar}.

The uplink transmission scenario is considered in which one cluster of two roadside sensors is handled by a single CE and BR. Both roadside sensors and BR communicate through single antennas. The forward channel link between CE and $k^{th}$ roadside sensor is denoted by $H_{f,k}$, while $H_{b,k}$  is used to denote the backscatter link between $k^{th}$ roadside sensor and BR, for $k \in \{1,2\}$. The channel coefficients of the forward and backscatter link are consist of the following components, respectively.

\begin{equation}
H_{f,k} = d^{-\alpha}_{f,k} \times h_{f,k},\label{eq 1}
\end{equation}
  and 
  \begin{equation}
  H_{b,k} = d^{-\alpha}_{b,k} \times h_{b,k},\label{eq 2}
  \end{equation}
where $h_{f,k}$ and $h_{b,k}$ represent fast fading component of forward and backscatter link, respectively. $d_{f,k}$ and  $d_{b,k}$ are the distance from CE to $k^{th}$ roadside sensor and $k^{th}$ roadside sensor to BR, respectively. $\alpha$ is used as path-loss exponent.
\begin{table}
	\fontsize{9}{11}\selectfont
	\centering
	\caption{THE LIST OF DIFFERENT SYMBOLS AND THEIR DEFINITIONS}
	\label{tabby}
	\begin{tabular}{l|l} 
		\hline 
		Symbol & Definition  \\
		\hline 
		$H_{f,k}$ & Forward  link channel coefficients of $k^{th}$ roadside \\ &  sensor.  \\
		$H_{b,k}$ & Backscatter link channel coefficients of $k^{th}$   \\ & roadside sensor.\\
		$h_{f,k}$ & Forward link fast fading component of $k^{th}$ \\ &  roadside sensor.\\
		$h_{b,k}$ & Backscatter link fast fading component of $k^{th}$ \\ & roadside sensor.\\
		$d_{f,k}$ & Distance from CE to $k^{th}$ roadside sensor.\\
		$d_{b,k}$ & Distance from $k^{th}$ roadside sensor to BR.\\
		$\alpha$ & Path loss exponent.\\
		$P^I_k$ & Power of incident RF signal at $k^{th}$ roadside \\ & sensor. \\
		$P_{ce}$ & CE transmit power. \\
		$P^{H_t}_k,P^{H_h}_k$ & Energy harvested by $k^{th}$ roadside sensor during\\& transmission mode and energy harvesting mode. \\
		$\xi$ & Efficiency of energy harvester. \\
		$\Gamma_{k}$ & Reflection coefficient of $K^{th}$ roadside sensor. \\
		$s_k$ & Reflected signal from $k^{th}$ roadside sensor \\ & towards BR.\\
		$x_k$ & Transmitted signal from $k^{th}$ roadside sensor \\ & towards BR.\\
		$n,\sigma^2_n$ & AWGN and its variance at RSU.\\
		$H_k$ & Overall channel link between $k^{th}$ roadside sensor \\ & and BR.\\
		$\hat{H}_{k}, \epsilon_k$ & Estimated channel gain and error of $k^{th}$ roadside \\ & sensor. \\
		$\sigma^{2}_{\hat{H}_{k}}, \sigma^{2}_{e_k}$ & Estimated channel and its error variance of $k^{th}$ \\ & roadside sensor.\\ 
		$y$ & Received signal at RSU with perfect CSI.\\
		$\hat{y}$ & Received signal at RSU with imperfect CSI.\\
		$R, \overline{R} $ & Sum-rate of roadside sensors before and after \\ & logrithmic approximation.\\
		$\gamma_{k}$ & SINR at $k^{th}$ roadside sensor.\\
		$P_T$ & Total power consumed by the cluster of sensors.\\
		$P^c_{ce}, P^c_{RSU}$ & CE and RSU circuit power. \\
		$P^c_{RS}$ & Roadside sensor circuit power. \\
		$\lambda,\boldsymbol{\mu}, \boldsymbol{\beta} $ & Lagrangian multipliers. \\
		$\varkappa_k$ & Power amplifier efficiency.\\
		$T_{t,k},T_{h,k}$ & Time coefficients of transmission and energy \\& harvesting mode of $k^{th}$ roadside sensor.\\ 
		\hline
	\end{tabular}
\end{table}
Let $P_{ce}$ is the power of the carrier signal emitted from CE. Then, the power of incident RF signal at $k^{th}$ roadside sensor is 
\begin{equation}
P^I_k = P_{ce}|H_{f,k}|^2. \label{eq 3}
\end{equation}
Roadside sensor can harvest energy from RF signals in both energy harvesting mode and transmission mode. Therefore, the energy harvested by $k^{th}$ roadside sensor during transmission mode ($T_t$) and energy harvesting mode ($T_h$) is given as follow
\begin{equation}
P^{H_{t}}_{k} = \xi(1-\Gamma_k) P^I_k T_{t,k}, \label{eq 4} 
\end{equation}
and 
\begin{equation}
 P^{H_{h}}_{k} = \xi P^I_k T_{h,k}, \label{eq 5}
\end{equation}
where $\xi$ denotes the efficiency of energy harvester while $\Gamma_k$ represents reflection coefficient of $k^{th}$ roadside sensor, where $0< \Gamma_k \le 1$. The reflected signal from $k^{th}$ roadside sensor toward BR installed on RSU is
\begin{equation}
s_k = \sqrt{P_{ce}\Gamma_k}H_{b,k}x_k , \label{eq 6}
\end{equation}
where $x_k$ is the transmitted signal form $k^{th}$ roadside sensor, satisfying $\mathbb{E}(|x_k|^2)=1$. Then , the signal received at RSU is given as
\begin{equation}
y=\sum_{k=1}^{2}H_{f,k}s_k+n,\label{eq 7}
\end{equation}
or
\begin{figure*}
	\centering
	\includegraphics[width=120mm]{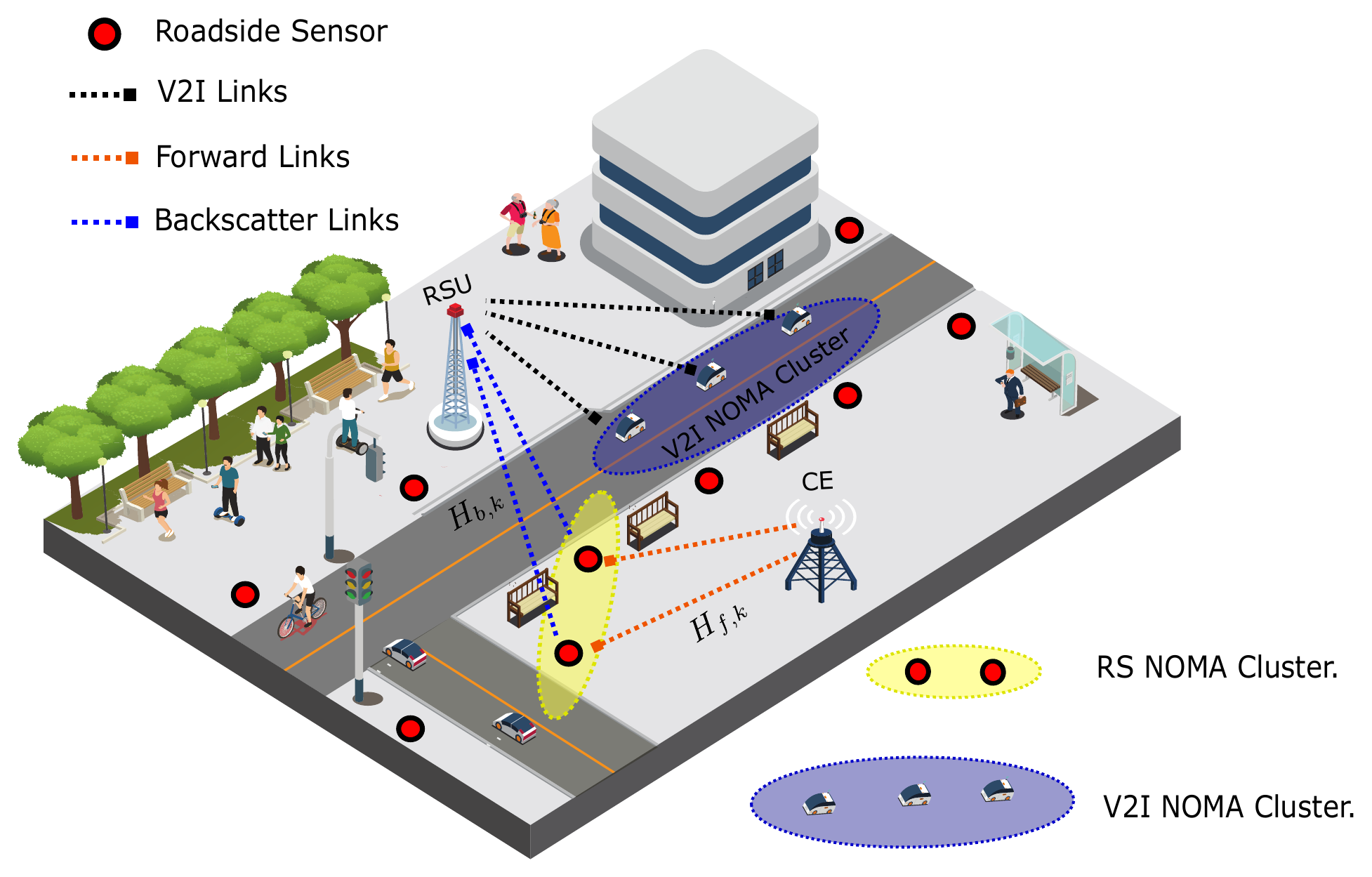}
	\caption{Illustration of System model}
	\label{FIG:1}
\end{figure*}
\begin{equation}
y = \sum_{k=1}^{2} \sqrt{P_{ce}\Gamma_k}H_{f,k}H_{b,k}x_k+n,\label{eq 8}
\end{equation}
where $H_k= H_{f,k}H_{b,k}$ is the overall channel link between $k^{th}$ roadside sensor and BR, which can be estimated through LMMSE method \cite{Y. Zhang}. $H_k$ can be written as $H_k= d_k \times h_k$ \cite{Y. Ye}, where $d_k= d^{-\alpha}_{b,k} \times d^{-\alpha}_{f,k} $ and $h_k = h_{b,k} \times h_{f,k}$. In practice, it is very challenging for backscatter aided communications to guarantee accurate CSI at all the time \cite{Y. Zhang}. Therefore, it is necessary to investigate the backscatter aided sensors to RSU communication with imperfect CSI. Under imperfect CSI estimation, $H_k$ can be expressed by using the minimum mean square error (MMSE) channel estimation error model \cite{A. Ihsan, Wang wang} as follow

 \begin{equation}
H_{k} = \hat{H}_{k}+\epsilon_k,\label{eq 9}
\end{equation}
where $\hat{H}_{k}\sim\mathcal{CN}(0,\,\sigma^{2}_{\hat{H}_{k}})$ is the estimated channel gain with variance $\sigma^{2}_{\hat{H}_{k}}=d_k - \sigma^{2}_{e_k}$ \cite{Z. Yang}, while $\epsilon_k\sim\mathcal{CN}(0,\,\sigma^{2}_{e_k})$ is the estimated channel error, which is Gaussian distributed with zero mean and variance $\sigma^{2}_{e_k}$. The relative channel error can be represented as $\rho_k = \frac {\sigma^{2}_{e_k}}{d_k}$. Then, $\sigma^{2}_{e_k}={\rho_k}{d_k}$ and $\sigma^{2}_{\hat{H}_{k}}= (1-{\rho_k})d_k$ \cite{J. Men}. For convenience of analysis, the case of constant estimation error $(\sigma^{2}_{e_k}=\sigma^{2}_{e})$ for all NOMA users is considered in this paper \cite{Z. Yang, M. R. Zamani}. 
Thus, the signal received at RSU under imperfect CSI is given as
\begin{equation}
\hat{y} = \sum_{k=1}^{2} \sqrt{P_{ce}\Gamma_k}\hat{H}_{k}x_k+\epsilon\sum_{k=1}^{2}\sqrt{P_{ce}\Gamma_k}x_k+n.\label{eq 10}
\end{equation}

Without loss of generality, roadside sensors channel gains are arranged as: $|\hat{H}_{2}|^2>|\hat{H}_{1}|^2$. Unlink downlink scenario, In the uplink, all information received at RSU from a cluster of roadside sensors are desired signals, although they give rise to multiuser interference. In downlink NOMA, SIC is implemented in ascending order i.e. the roadside sensor with lower channel gain is first decoded and removed. While in the uplink, the RSU can decode the roadside sensors signals in an arbitrary order, as all information is desired
at the RSU \cite{M. Zeng}. However, despite that, to implement SIC and decode the received information at the RSU, tuning the value of reflection coefficient of each roadside sensor to different values should be exploited in such a way that it maintained the distinctness among various signals. For the 
analysis, we assume that the SIC order that decodes the information of the 2nd roadside sensor (highest channel gain) first is employed at the RSU throughout this work \cite{M. S. Ali}. According to the NOMA principles, the sum-rate of roadside sensors is given by

\begin{align}
R= &BWT_{t,1}\log_2\bigg(1+\gamma_{1}\bigg)+BWT_{t,2}\log_2\bigg(1+\gamma_{2}\bigg), \label{eq 11}
\end{align}
where
\begin{align}
\gamma_{1}=\bigg(\frac{P_{ce}\Gamma_1|\hat{H}_{1}|^2}{\sigma^{2}_{e} P_{ce}\sum_{k=1}^{2}\Gamma_k+\sigma^2_n}\bigg), \label{eq 12}
\end{align}
and
\begin{align}
\gamma_{2}=\bigg(\frac{P_{ce}\Gamma_2|\hat{H}_{2}|^2}{{P_{ce}\Gamma_1|\hat{H}_{1}|^2}+\sigma^{2}_{e} P_{ce}\sum_{k=1}^{2}\Gamma_k+\sigma^2_n}\bigg). \label{eq 13}
\end{align}

The total power consumed by the cluster is the circuit power consumption of the roadside sensors in backscatter mode. The circuit operation is performed through the harvested power from the carrier signal during transmission mode as well as in energy harvesting mode. The power consumed by the cluster can be obtained by the sum of the harvested powers of roadside sensors paired in that cluster. Besides it, total energy consumption also consists of the energy consumed by the CE to transmit RF signals and RSU to receive the signals of roadside sensors. Therefore, the total power consumption of the system is as follow
\begin{align}
P_T= \sum_{k=1}^{2}\frac {P_{ce}}{\varkappa_k}\big(T_{t,k}+T_{h,k}\big)+P^c_{ce}+P^c_{RSU}, \label{eq 14}
\end{align}
where $\varkappa_k$ is power amplifier efficiency that ranges between zero and one while $P^c_{ce}$ and $P^c_{RSU}$ are the constant circuit power consumed by CE and RSU respectively. The energy efficiency of the proposed system model is defined as the ratio of the sum rate of roadside sensor to the consumed power of system in bits per Hertz per Joule (bits/Hz/J) \cite{Q. W Chen 1, Q. W Chen 2} as follows

\begin{align}
EE = \frac{R}{P_T}. \label{eq 15}
\end{align}

\section{Problem Formulation}
Energy-efficient NOMA-enabled wireless powered passive sensors to infrastructure communication under imperfect CSI is the primary objective of the optimization problem. The energy efficiency maximization problem is formulated as
\begin{align}
&\underset { P_{ce},\boldsymbol{\Gamma}}{\max}  {EE}= \underset { P_{ce},\boldsymbol{\Gamma}}{\max} \frac {R} {P_T}, \nonumber\\
s.t. \,
& C1: {P_{ce}\Gamma_1|\hat{H}_{1}|^2} \geq (2^{\frac {R_{min}}{T_{t,1}}}-1)\bigg({\sigma^{2}_{e} P_{ce}\sum_{k=1}^{2}\Gamma_k+\sigma^2_n}\bigg),\nonumber\\
& C2: {P_{ce}\Gamma_2|\hat{H}_{2}|^2} \geq (2^{\frac {R_{min}}{T_{t,2}}}-1)\bigg({P_{ce}\Gamma_1|\hat{H}_{1}|^2}\nonumber\\
&+{\sigma^{2}_{e} P_{ce}\sum_{k=1}^{2}\Gamma_k+\sigma^2_n}\bigg),\nonumber\\
&C3: 0 \leq P_{ce} \leq P_{max},\nonumber \\
&C4: 0 < \Gamma_k \le 1,  k \in \{1,2\},\nonumber  \\
&C5: P^{H_h}_{k}+P^{H_t}_{k}\geq P^c_{RS}T_{t,k},  k \in \{1,2\}, \label{eq:16}
\end{align}
where $EE$ represents the energy efficiency of the wireless powered sensor communications in NOMA-enabled ITS. $\boldsymbol{\Gamma} = \{\Gamma_{1}, \Gamma_{2} \}$ is the vector of reflection coefficient of roadside sensors. C1 and C2 enforce roadside sensors to backscatter their information towards RSU with minimum QoS (minimum required rate) requirement. C3 ensures that the CE transmit power ($P_{ce}$) cannot exceed its maximum transmit power ($P_{max}$). C4 limits the reflection coefficient of sensors between 0 and 1. C5 guarantees that the energy harvested by the $k^{th}$ sensor should exceed their minimum required circuit power ($P^c_{RS}$).

The objective function in the problem defined in Eq. (\ref{eq:16}) has a non-linear fractional form, which is challenging to solve. The logarithmic approximation is implemented \cite{Papandriopoulos}, which reduces complexity and transforms the optimization problem into tractable concave-convex fractional programming (CCFP) problem. The approximation is done as follow
\begin{equation}
\Pi\log_{2}(z)+\Phi \leq \log_{2}(1+z), \label{eq:17}
\end{equation}
for any $z\geq 0$ , where $ \Pi = \frac {z_0}{1+z_0} $ and $\Phi=\log_{2}(1+{z_0})-\frac {z_0}{1+z_0}\log_{2}(z_0) $. When $z =z_0$, the bound becomes tight. Through lower bound of inequality in Eq. (\ref{eq:17}), the sum-rate of roadside sensors is presented as
\begin{equation}
\overline{R}= \sum_{k=1}^{2}BWT_{t,k}(\Pi_{k}\log_{2}({ \gamma_{k}})+\Phi_{k}), \label{eq:18}
\end{equation}
where
\begin{equation}
\Pi_{k} = \frac {{\gamma_{k}}}{1+{\gamma_{k}}}, \label{eq:19}
\end{equation}
and
\begin{equation}
\Phi_{k}=\log_{2}(1+{\gamma_{k}})-\frac {{ \gamma_{k}}}{1+{ \gamma_{k}}}\log_{2}({\gamma_{k}}). \label{eq:20}
\end{equation}
Hence the updated optimization problem can be formulated as
\begin{equation}
\begin{split}
&\underset { P_{ce},\boldsymbol{\Gamma}}{\max}  {EE}= \underset { P_{ce},\boldsymbol{\Gamma}}{\max} \frac {\overline{R}} {P_T}, \\
s.t. \,
& C1: {P_{ce}\Gamma_1|\hat{H}_{1}|^2} \geq \bigg(2^{\frac{R_{min}-T_{t,1}\Phi_1}{T_{t,1}\Pi_1}}\bigg)\bigg({\sigma^{2}_{e} P_{ce}\sum_{k=1}^{2}\Gamma_k+\sigma^2_n}\bigg),\\
& C2: {P_{ce}\Gamma_2|\hat{H}_{2}|^2} \geq \bigg(2^{\frac{R_{min}-T_{t,2}\Phi_2}{T_{t,2}\Pi_2}}\bigg)\bigg({P_{ce}\Gamma_1|\hat{H}_{1}|^2}\\
&+{\sigma^{2}_{e} P_{ce}\sum_{k=1}^{2}\Gamma_k+\sigma^2_n}\bigg),\\
&C3: 0 \leq P_{ce} \leq P_{max}, \\
&C4: 0 < \Gamma_k \le 1,  k \in \{1,2\},  \\
&C5: P^{H_h}_{k}+P^{H_t}_{k}\geq P^c_{RS}T_{t,k}. \label{eq:21}
\end{split}
\end{equation}
The EE maximization problem is coupled on two kinds of optimization variables i.e., $P_{ce}$, and $\boldsymbol{\Gamma}$ in problem defined in Eq. (\ref{eq:21}). Thus, it is very hard to find a globally optimal solution directly. It demands an approximate optimal algorithm through alternating optimization algorithm. Therefore, this problem can be solved through alternating optimization algorithm in two stages: $i)$ In first stage, on the fixed value of reflection coefficient of roadside sensors, we compute the energy efficient transmit power
of the carrier emitter $P_{ce}$; $ii)$ we substitute the optimal $P^*_{ce}$ obtained in stage 1 into the original problem and optimize the reflection coefficients of the roadside sensors in Stage 2.

\subsection{Energy Efficent transmit power allocation for CE}
Given the reflection coefficients of roadside sensors, the optimization problem in Eq. (\ref{eq:21}) can be simplified to CE transmit power optimization as follow
\begin{equation}
\begin{split}
&\underset { P_{ce}}{\max}  {EE}= \underset { P_{ce}}{\max} \frac {\overline{R}} {P_T}, \\
s.t. \,
& C1-C3, C5.
\label{eq:22}
\end{split}
\end{equation}
By using the following lemma, we demonstrate
that $\overline{R}$ is a concave function with respect to $P_{ce}$.\\
\\
$\mathit{Lemma} \: 1:$
\begin{align}
\overline{R}= &BWT_{t,1}(\Pi_{1}\log_{2}\bigg(\frac{P_{ce}\Gamma_1|\hat{H}_{1}|^2}{\underbrace{\sigma^{2}_{e} P_{ce}\sum_{k=1}^{2}\Gamma_k+\sigma^2_n}_\text{X}}\bigg)+\Phi_{1})+\\
&BWT_{t,2}(\Pi_{2}\log_{2}\bigg(\frac{P_{ce}\Gamma_2|\hat{H}_{2}|^2}{\underbrace{{P_{ce}\Gamma_1|\hat{H}_{1}|^2}+\sigma^{2}_{e} P_{ce}\sum_{k=1}^{2}\Gamma_k+\sigma^2_n}_\text{Y}}\bigg)+\Phi_{2}) \nonumber \label{eq 23}
\end{align}
is a concave function with respect to $P_{ce}$.\\
\\
$\mathit{Proof:}$ Please, refer to Appendix A.\\

As, $\overline{R}$ is concave with respect to $P_{ce}$ and the objective function denominator is an affine function of $P_{ce}$ in Eq. (\ref{eq:22}). Thus, the problem in Eq. (\ref{eq:22}) is in the form of a concave-convex fractional programming (CCFP) problem. Which can be solved efficiently through Dinkelbach’s algorithm \cite{Shen}. By using Dinkelbach method, problem in Eq. (\ref{eq:22}) can be transformed as 
\begin{equation}
\begin{split}
&\underset { P_{ce}}{\max}  {EE} =\underset { P_{ce}}{\max} F(\psi)=\underset { P_{ce}}{\max} {\overline{R}} -\psi{P_T}, \\
s.t. \,
& C1-C3, C5, \label{eq:24}
\end{split}
\end{equation}
where $\psi$ is the real parameter. Computing the roots of $F(\psi)$ is analogous to solving the objective function in Eq. (\ref{eq:22}) \cite{M. R. Zamani}. $F(\psi)$ is negative when $\psi$ approaches infinity,
while $F(\psi)$ is positive when $\psi$ approaches minus infinity. $F(\psi)$ is convex with respect to $\psi$. The convex problem in Eq. (\ref{eq:24}) is solved by employing the Lagrangian dual decomposition method. The Lagrangian function is presented in Eq. (\ref{eq:25}),

\begin{table*}
\begin{align}
&\mathit{L(P_{ce},\boldsymbol{\mu},\lambda,\boldsymbol{\beta})}=  \nonumber \\
&BWT_{t,1}\Bigg(\Pi_{1}\log_{2}\bigg(\frac{P_{ce}\Gamma_1|\hat{H}_{1}|^2}{\sigma^{2}_{e} P_{ce}\sum_{k=1}^{2}\Gamma_k+\sigma^2_n}\bigg)+\Phi_{1}\Bigg)+BWT_{t,2}\Bigg(\Pi_{2}\log_{2}\bigg(\frac{P_{ce}\Gamma_2|\hat{H}_{2}|^2}{{P_{ce}\Gamma_1|\hat{H}_{1}|^2}+\sigma^{2}_{e} P_{ce}\sum_{k=1}^{2}\Gamma_k+\sigma^2_n}\bigg)+\Phi_{2}\Bigg)-\psi\Bigg(\sum_{k=1}^{2}\frac {P_{ce}}{\varkappa_k}\big(T_{t,k}+T_{h,k}\big)+P^c_{ce} \nonumber\\
&+P^c_{RSU}\Bigg)+\mu_1\Bigg({P_{ce}\Gamma_1|\hat{H}_{1}|^2}-\Big(2^{{\frac{R_{min}-T_{t,1}\Phi_1}{T_{t,1}\Pi_1}}}\Big)\Big({\sum_{k=1}^{2}\sigma^{2}_{e}P_{ce}\Gamma_k+\sigma^2_n}\Big)\Bigg)+\mu_2\Bigg({P_{ce}\Gamma_2|\hat{H}_{2}|^2}-\Big(2^{\frac{R_{min}-T_{t,2}\Phi_2}{T_{t,2}\Pi_2}}\Big)\Big({P_{ce}\Gamma_1|\hat{H}_{1}|^2}+{\sum_{k=1}^{2}\sigma^{2}_{e}P_{ce}\Gamma_k+\sigma^2_n}\Big)\Bigg) \nonumber\\
&+\lambda\Bigg(P_{max}-P_{ce}\Bigg)+\beta_1\Bigg(\xi(1-\Gamma_1)P_{ce}|H_{f,1}|^2T_{t,1}+\xi P_{ce}|H_{f,1}|^2T_{h,1}-P^c_{RS}T_{t,1}\Bigg)+\beta_2\Bigg(\xi(1-\Gamma_2)P_{ce}|H_{f,2}|^2T_{t,2}+\xi P_{ce}|H_{f,2}|^2T_{h,2} -P^c_{RS}T_{h,2}\Bigg), \label{eq:25}
\end{align}
\hrule\vspace{5mm}
\end{table*} 
where $\boldsymbol{\mu}=\{\mu_1,\mu_2 \}$, $\boldsymbol{\beta}=\{\beta_1,\beta_2 \}$, and $\lambda$ are the Lagrange multipliers. Constraints are KKT conditions for optimizing the power allocation for CE.
\\
\\
$\mathit{Lemma} \: 2:$
\\
\\
The closed-form solution of optimal $P_{ce}$ can be expressed as
\begin{align}
P^*_{ce}= \sqrt[3]{q+\sqrt{q^2+\big(r-p^2\big)^3}}+\sqrt[3]{q-\sqrt{q^2+\big(r-p^2\big)^3}}+p \label{eq:26}
\end{align}
$\mathit{Proof:}$ Please, refer to Appendix B \\

 Given the optimal CE transmit power allocation policy in Eq. (\ref{eq:26}), the primal problem's Lagrangian multipliers can be determined and updated iteratively by employing the sub-gradient method as presented in Eq. (\ref{eq:27}) - Eq. (\ref{eq:31}).
 \begin{table*}
\begin{align}
\lambda(iter+1)=&\Big[\lambda(iter)-\omega_1(iter)\Big(P_{max}-P_{ce}\Big)\Big]^+, \label{eq:27} \\
\mu_{1}(iter+1)=&\Big[\mu_{1}(iter)-\omega_2(iter)\Big({P_{ce}\Gamma_1|\hat{H}_{1}|^2}-\Big(2^{\frac{R_{min}-T_{t,1}\Phi_1}{T_{t,1}\Pi_1}}\Big)
\Big({\sum_{k=1}^{2}\sigma^{2}_{e}P_{ce}\Gamma_k+\sigma^2_n}\Big)\Big)\Big]^+, \label{eq:28} \\
\mu_{2}(iter+1)=&\Big[\mu_{2}(iter)-\omega_3(iter)\Big({P_{ce}\Gamma_2|\hat{H}_{2}|^2}-\Big(2^{\frac{R_{min}-T_{t,2}\Phi_2}{T_{t,2}\Pi_2}}\Big)\Big({{P_{ce}\Gamma_1|\hat{H}_{1}|^2}+\sum_{k=1}^{2}\sigma^{2}_{e} P_{ce}\Gamma_k+\sigma^2_n}\Big)\Big)\Big]^+,  \label{eq:29} \\
\beta_1(iter+1)=&\Big[\beta_1(iter)-\omega_4(iter)\Big(\xi(1-\Gamma_1)P_{ce}|H_{f,1}|^2 T_{t,1}+\xi P_{ce}|H_{f,1}|^2T_{h,1}-P^c_{RS}T_{t,1}\Big)\Big]^+,  \label{eq:30} \\
\beta_2(iter+1)=&\Big[\beta_2(iter)-\omega_5(iter)\Big(\xi(1-\Gamma_2)P_{ce}|H_{f,2}|^2 T_{t,2}+\xi P_{ce}|H_{f,2}|^2T_{h,2}-P^c_{RS}T_{t,2}\Big)\Big]^+,  \label{eq:31}
\end{align}
\hrule\vspace{5mm}
\end{table*}
\\
Where $iter$ is used for iteration index. $\omega_1$, $\omega_2$,$\omega_3$, $\omega_4$, and $\omega_5$  present positive step sizes. The appropriate step sizes should be used for the convergence to an optimal solution.

\subsection{Efficient selection of reflection coefficient for roadside sensor}
Through simplified Cardano's formulae, a closed-form of optimal power for CE is obtained. Now the optimization problem for allocating efficient reflection coefficients to roadside sensors under their QoS and required circuit power constraint can be rewritten as,
\begin{align}
		& \underset { \boldsymbol{\Gamma}}{\max} \: {\overline R} -\psi{P_T}, \nonumber\\
		s.t. \,
		& C1: {P^*_{ce}\Gamma_1|\hat{H}_{1}|^2} \geq \bigg(2^{\frac{R_{min}-T_{t,1}\Phi_1}{T_{t,1}\Pi_1}}\bigg)\bigg({\sum_{k=1}^{2}\sigma^{2}_{e}P^*_{ce}\Gamma_k+\sigma^2_n}\bigg),\nonumber\\
		& C2: {P^*_{ce}\Gamma_2|\hat{H}_{2}|^2} \geq \bigg(2^{\frac{R_{min}-\Phi_2}{\Pi_2}}\bigg)\bigg({P^*_{ce}\Gamma_1|\hat{H}_{2}|^2}\nonumber\\
		&+{\sum_{k=1}^{2}\sigma^{2}_{e}P^*_{ce}\Gamma_k+\sigma^2_n}\bigg),\nonumber\\
		&C3: 0 < \Gamma_k \le 1,  k \in \{1,2\}, \nonumber \\
		&C4: \sum_{k=1}^{2}\Gamma_k = \theta ,\nonumber  \\
		&C5: P^{H_h}_{k}+P^{H_t}_{k}\geq P^c_{RS}T_{t,k}. \label{eq:32}
\end{align}
To solve optimization problem defined in Eq. (\ref{eq:32}), first we will present in Lemma 2 that $\overline{R}$ is a  concave function with respect to $\Gamma_{1}$ and  $\Gamma_{2}$.\\
\\
$\mathit{Lemma 3:}$
\begin{align}
\overline{R}= &BWT_{t,1}(\Pi_{1}\log_{2}\bigg(\frac{P^*_{ce}\Gamma_1|\hat{H}_{1}|^2}{\sigma^{2}_{e} P^*_{ce}\theta+\sigma^2_n}\bigg)+\Phi_{1})+ \label{eq 33}\\
&BWT_{t,2}(\Pi_{2}\log_{2}\bigg(\frac{P^*_{ce}\Gamma_2|\hat{H}_{2}|^2}{{P^*_{ce}\Gamma_1|\hat{H}_{1}|^2}+\sigma^{2}_{e} P^*_{ce}\theta+\sigma^2_n}\bigg)+\Phi_{2}) \nonumber
\end{align}
is a concave function with respect to $\Gamma_{1}$ and  $\Gamma_{2}$.
\\
\\
$\mathit{Proof:}$ Please, refer to Appendix C\\
\\
As, $\overline{R}$ is concave with respect to $\Gamma_{1}$ and  $\Gamma_{2}$. Therefore maximum value of optimization problem defined in Eq. (\ref{eq:32}) can be obtained by the upper bound of reflection coefficient of roadside sensor as follow,\\
\\
$\mathit{Lemma \: 4:}$
\\
\\
The optimal reflection coefficient of the optimization problem defined in Eq. (\ref{eq:32}) is presented as
 
\begin{align}
&\Gamma_{k}^*=  max\Bigg\{\frac {2^{\aleph_k}(\sigma^{2}_{T}+I_{NOMA})}{P_{ce}^*|\hat{H}_{k}|^2}, min \Big\{1- \frac{P^c_{RS}}{P^I_k}+\frac{T_{h,k}}{T_{t,k}}, 1\Big\}\Bigg\}  \label{eq 34} 
\end{align}
where, $\aleph_k=\frac{R_{min}-T_{t,k}\Phi_1}{T_{t,k}\Pi_k}$,  $\sigma^{2}_{T}=\sigma^{2}_{e} P_{ce}^*\theta+\sigma^2_n$ and $I_{NOMA}=\sum_{l=1}^{k-1}P^*_{ce}\Gamma_l|\hat{H}_{l}|^2$\\
\\
$\mathit{Proof:}$ 
\\
For the given optimal $P_{ce}^*$, the objective function of optimization problem in Eq. (\ref{eq:32}) increases with increase of $\boldsymbol{\Gamma}$. Therefore, the optimal reflection coefficient of roadside sensor can be computed by the upper bound of reflection coefficient. The range of $\Gamma_{k}$ from the optimization problem in Eq. (\ref{eq:32}) can be determined by combining constraints C1, C3, and C5 for $\Gamma_{1}$ and constraints C2, C3, and C5 for $\Gamma_{2}$ .  After some simple mathematical computations, the range of $\Gamma_{k}$ can be presented as

 \begin{align}
   \frac {2^{\aleph_k}(\sigma^{2}_{T}+I_{NOMA})}{P_{ce}^*|\hat{H}_{k}|^2} \le \Gamma_{k} \le min \Big\{1- \frac{P^c_{RS}}{P^I_k}+\frac{T_{h,k}}{T_{t,k}},1\Big\} \label{eq 35}  
 \end{align}  
Thus, optimal $\Gamma_{k}^*$ can be calculated as $max\Bigg\{\frac {2^{\aleph_k}(\sigma^{2}_{T}+I_{NOMA})}{P_{ce}^*|\hat{H}_{k}|^2}, min \Big\{1- \frac{P^c_{RS}}{P^I_k}+\frac{T_{h,k}}{T_{t,k}},1\Big\}\Bigg\}$.
\\
Through Lemma 4, optimal reflection coefficients of roadside sensors can be determined. It is allocated to the roadside sensors in the following manner,
\begin{itemize}
	\item When the condition $\frac {2^{\aleph_k}(\sigma^{2}_{T}+I_{NOMA})}{P_{ce}^*|\hat{H}_{k}|^2} \le 1- \frac{P^c_{RS}}{P^I_k}+\frac{T_{h,k}}{T_{t,k}} \le 1 $ is satisfied. Then, Lemma 4 will results in  $\Gamma_{k}^*= 1- \frac{P^c_{RS}}{P^I_k}+\frac{T_{h,k}}{T_{t,k}} $. This condition implies that the quality of service constraint and and circuit power constraint of sensors is guaranteed simulataneously. In such condition,  $\Gamma_{k}^*= 1- \frac{P^c_{RS}}{P^I_k}+\frac{T_{h,k}}{T_{t,k}} $ presents the portion of backscattering signal in received RF signal at roadside sensor during transmission mode $T_t$ while remaining portion of received RF signal is used for harvesting.
	\item When the condition $\frac {2^{\aleph_k}(\sigma^{2}_{T}+I_{NOMA})}{P_{ce}^*|\hat{H}_{k}|^2} \le 1 \le 1- \frac{P^c_{RS}}{P^I_k}+\frac{T_{h,k}}{T_{t,k}}  $ is satisfied. Then, it implies that the energy harvested during energy harvesting mode $T_h$ is enough for the circuit operation of roadside sensor and in transmission mode $T_t$ all the received RF signal at roadside sensor can be used for backscattering for EE maximization. In this condition Lemma 4 results in $\Gamma_{k}^*= 1$. In this condition, if both sensors which are backscattering using NOMA results in reflection coefficient of 1, then the reflection coefficient of near sensor should be calculated as 1-$\nu$. The value of $\nu$ can be selected in such a way to satisfy ${P^*_{ce}\Gamma_2|\hat{H}_{2}|^2}-{P^*_{ce}\Gamma_1|\hat{H}_{1}|^2} \leq P_{gap}$ (SIC constraint) \cite{M. S. Ali}.
	\item When the condition $\frac {2^{\aleph_k}(\sigma^{2}_{T}+I_{NOMA})}{P_{ce}^*|\hat{H}_{k}|^2} > 1- \frac{P^c_{s}}{P^I_k}+\frac{T_{h,k}}{T_{t,k}}  $ is satisfied. Then, it implies that QoS constraint and circuit power constraint of roadside sensor can not be guarnteed simulataneously. In such case, optimization problem defined in Eq. (\ref{eq:32}) is infeasible. \\
	The details of proposed algorithm AOBWS is summarized in Algorithm 1.    
\end{itemize}
Once RSU received the information from roadside sensors, then that information can be provided to the vehicles through energy efficient schemes proposed in \cite{A. Ihsan, Khan W}.
\begin{algorithm}
	\caption{Alternating optimization for backscatter aided wireless powered sensors (AOBWS) algorithm }
	\label{pseudoPSO}
	\begin{algorithmic}[1]
		\State \textbf{Stage 1: OCETP}
		\State \textbf{Initialization:} Reflection coefficient of roadside sensor, maximum iterations $I_{max}$, and maximum tolerance $\delta_{max}$. Initialize the stepsizes and the dual variables($\boldsymbol{\beta}$, $\boldsymbol{\mu}$ and $\lambda$) and iteration index I = 1.
		\While {$I \leq I_{max} \quad \textbf {or} \quad|{\overline{R}}(I) - \psi(I)P_T| \geq \delta_{max}$}
		\State Compute ${\overline{R}}(I)$ by using  Eq. (\ref{eq:18})
		\State Compute $\psi(I)= \frac {\overline{R}(I)} {P_T}$
		\State Update dual variables $\lambda(I)$ and $\boldsymbol{\mu} (I)$ and $\boldsymbol{\beta}(I)$ by using Eq. (\ref{eq:27}), (\ref{eq:28}), (\ref{eq:29}), (\ref{eq:30}), and (\ref{eq:31}), respectively.
		\State Update the transmit power ${P_{ce}}(I+1)$ of CE by using equation (\ref{eq:26}) in \textbf{ Lemma: 2}.
		\State $ I = I+1$.
		\EndWhile\\
		\textbf{Output:} Optimal CE transmit power $P^*_{ce}$.
		\State \textbf{Stage 2: Optimal Reflection Coefficient }
		\State Under optimal $P^*_{ce}$ obtained at stage 1, compute optimal reflection coefficient of roadside sensor through \textbf{ Lemma: 4}.
		\State \textbf{Output:} Optimal $\boldsymbol{\Gamma^*}$ = \{$ \Gamma_{1}^*$, $\Gamma_{2}^*$\}
		\State \textbf{Algorithm 1 Output:} AOBWS = OCETP + Optimal reflection coefficient = $P^*_{ce}$ and $\boldsymbol{\Gamma^*}$. 
	\end{algorithmic}
\end{algorithm}
 \subsection{Complexity Analysis:} 
 This subsection presents the complexity analysis of the proposed AOBWS algorithm and ES algorithm as benchmark algorithm. In literature, ES algorithm is employed for NOMA optimal power allocation \cite{W. Saetan}. Moreover, it is also used as a benchmark algorithm \cite{Yu,Y. Liu,Cui}, because it shows the best performance as compare to all other algorithms but at high computational complexity. In this article, ES algorithm is used as a benchmark because it obtains global optimal EE performance for the considered system model. It searches over all possible search points in the search regions of CE transmit power and reflection coefficients of roadside sensor. However, it demands a large amount of computation as follows. If $P_{max}$ is the maximum transmit power of CE and $P_{step}$ is the step size for transmit power of CE. Then, there are $(\frac{P_{max}}{P_{step}})$ choices for the values of transmit power of CE. Similarly, if $\Gamma_{max}$ is the maximum value of reflection coefficient that can be assigned to the roadside sensor and $\Gamma_{step}$ is the step size for reflection coefficient, then there are $(\frac{\Gamma_{max}}{\Gamma_{step}})^K$ choices for the values of reflection coefficients of K sensors linked with BR installed on RSU.  Therefore, the complexity of ES algorithm is $\mathcal{O}(\frac{P_{max}}{P_{step}})+\mathcal{O}(\frac{\Gamma_{max}}{\Gamma_{step}})^{K}$. It can be noticed that the ES algorithm is computationally expensive but is used as a benchmark because of its global optimal solution. For practical implementation AOBWS algorithm is proposed, which requires the complexity order of $\mathcal{O}(IK^2+K)$.
 
AOBWS algorithm employs a two-stage procedure, in which the first stage is consists of an iterative algorithm (OCETP) and 2nd stage is the non-iterative algorithm. OCETP algorithm during each iteration requires $K$ operations to calculate EE. Where $K$ is the total number of roadside sensors linked using NOMA with BR installed on RSU. Furthermore, $K$ operations are required to update dual variables. If I is the number of iterations that the OCETP algorithm needs to converge, then the total complexity of the OCETP algorithm is $\mathcal{O}(IK^2)$. While in the 2nd Stage, AOBWS employs a non-iterative algorithm for optimal reflection coefficients, which requires K operations. After two-stage procedures, the complexity order of AOBWS becomes $\mathcal{O}(IK^2+K)$. So, the proposed AOBWS algorithm is always less computationally complex as compared to the ES algorithm. When the number of $K$ roadside sensors connected through NOMA with each RSU increases, the complexity of the ES algorithm increases exponentially while the proposed AOBWS algorithm still provides an optimal solution in polynomial time. The complexity analysis of analyzed algorithms is presented in Table III. 
\begin{table}
	\small
	\centering
	\caption{Complexity analysis of algorithms}\label{tbl2}
	\begin{tabular}{ |p{4cm}|p{4cm}| }
		\hline
		\textbf{Algorithm} & \textbf{Complexity} \\
		\hline
		Proposed AOBWS algorithm & $\mathcal{O}(IK^2+K)$  \\
		\hline
		ES algorithm (Benchmark) & $\mathcal{O}(\frac{P_{max}}{P_{step}})+\mathcal{O}(\frac{\Gamma_{max}}{\Gamma_{step}})^{K}$  \\
		\hline
	\end{tabular}
\end{table} 



\section{Simulations}
\begin{figure}
	\centering
	\includegraphics[width=92mm]{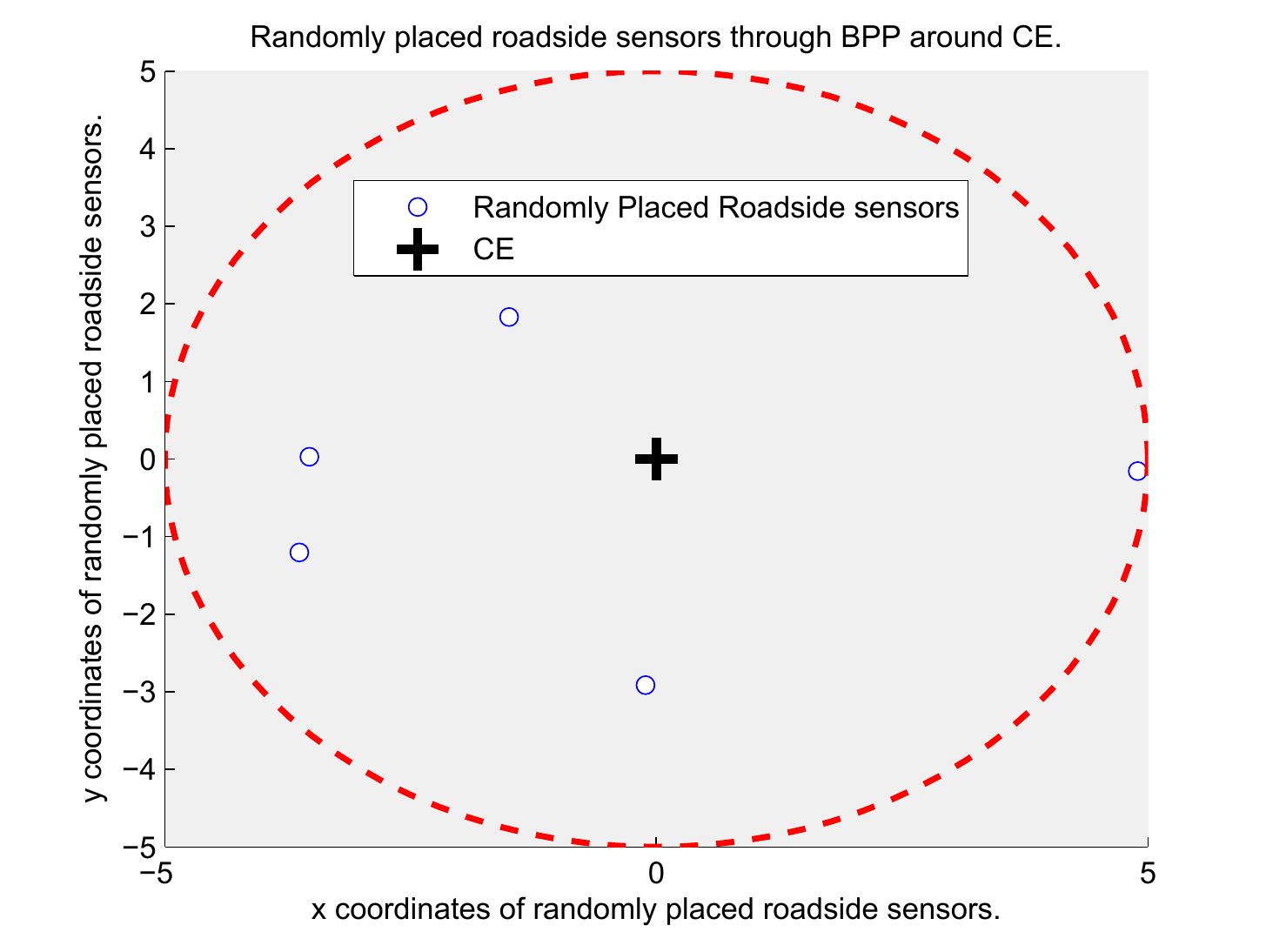}
	\caption{Random distribution of roadside sensors around CE through BPP}
	\label{FIG:2}
\end{figure}

\begin{table}
	\begin{center}
		\caption{Simulation Parameters.}\label{tbl2}
		\begin{tabular}{ |c|c|c| }
			\hline
			\textbf{Parameter} & \textbf{Value} \\
			\hline
			Bandwidth (BW) & 1 MHz  \\
			\hline
			CE Radius & 5 m  \\
			\hline
			Roadside sensors distribution around CE & BPP  \\
			\hline
			Noise power $(\sigma^2)$ & -114 dBm \\
			\hline
			Transmit power of CE $(P_{ce})$ & 0 dBm - 40 dBm \\
			\hline
			Relative channel errors $(\rho)$ & 0.001 - 0.009 \\
			\hline
			EH efficiency coefficient $(\xi)$ & 0.6 \\
			\hline
			power amplifier efficiency $(\varkappa)$ & $(0,1]$ \\
			\hline
			CE circuit power consumption $(P^c_{ce})$ & 100 mW \\
			\hline
			RSU circuit power consumption $(P^c_{RSU})$ & 1 W \\
			\hline
			Roadside sensor circuit power consumption $(P^c_s)$ & -35 dBm \\
			\hline
			Roadside sensor minimum data rate $R_{min}$ & 0.5 bps/Hz \\
			\hline
			Path loss & distance dependent.\\
			\hline
			path-loss exponent$(\alpha)$ & 4 \\ 
			\hline
			Fast fading & Rayleigh fading\\
			\hline
			
		\end{tabular}
	\end{center}
\end{table} 
This section presents the simulation results to evaluate the efficacy of the proposed algorithm in terms of energy efficiency and complexity. For this purpose, the proposed AOBWS algorithm is compared with the global optimal ES algorithm (benchmark algorithm). ES algorithm achieves global optimal solution but at the cost of high computational complexity and is impractical. It is observed the proposed AOBWS algorithm achieves very close performance to the global optimal ES with very low complexity and is therefore suitable for practical implementation. During simulations, roadside sensors are deployed randomly around the CE on the road and their locations are modeled as binomial point process (BPP) unless otherwise specified. In our simulations, each subchannel is serving two roadside sensors through
NOMA, which are selected randomly from the generated
sesnors through BPP as shown in Fig.~\ref{FIG:2}. The distance between far roadside sensor to RSU and near roadside sensor to RSU is set as 50 meters and 30 meters respectively unless otherwise specified. Time coefficient for roadside sensor transmission mode and energy harvesting mode is set to $T_{t,1} = T_{t,2} = 0.5$ and $T_{h,1} = T_{h,2} = 0.5$. Other main simulation parameters for our considered system model are described in Table IV.

\begin{figure}
	\centering
	\includegraphics[width=85mm]{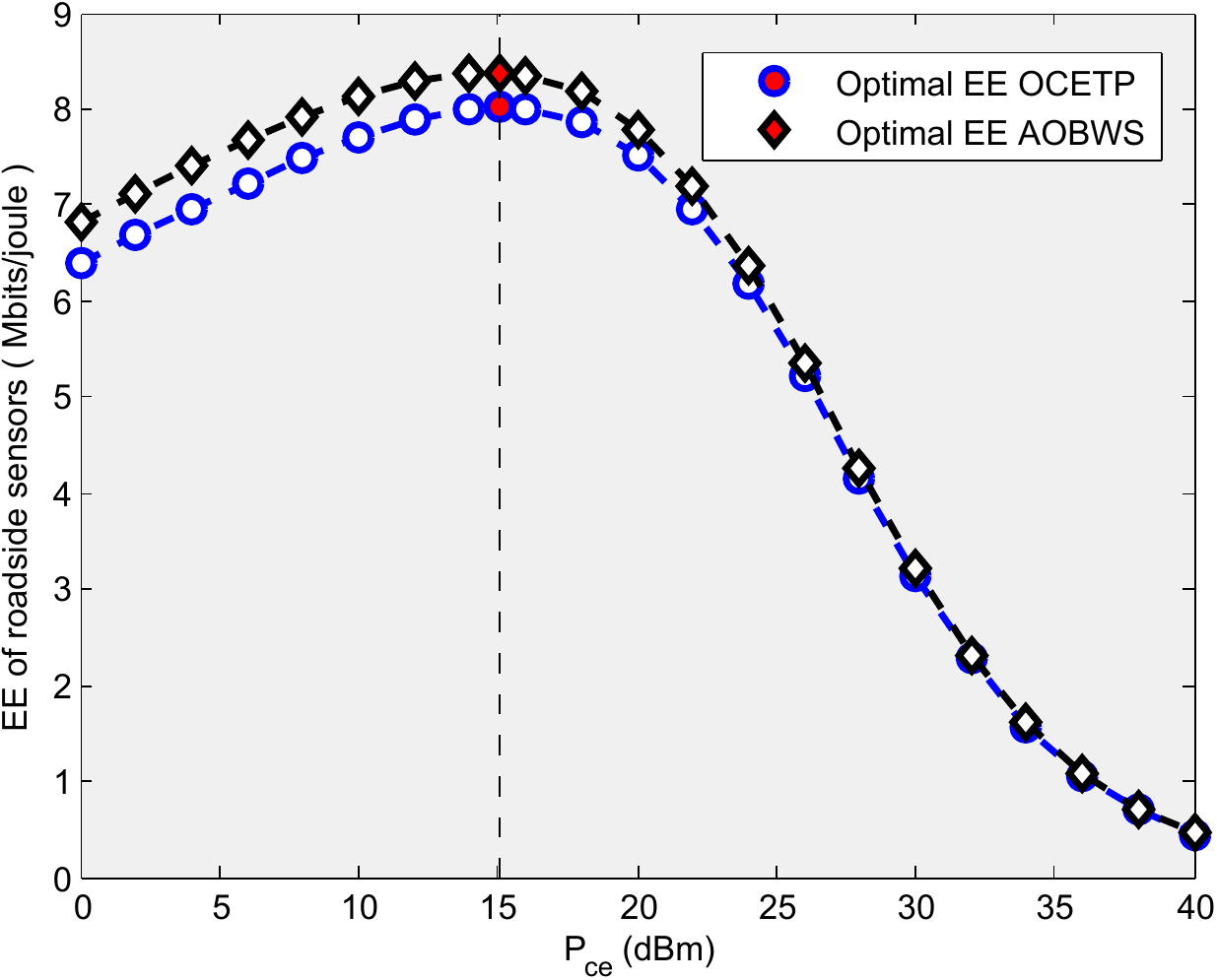}
	\caption{Optimal EE of roadside sensors achieved through proposed OCETP and AOBWS.}
	\label{FIG:3}
\end{figure}

Fig.~\ref{FIG:3} presents the EE of roadside sensors versus the transmit power of CE in dBm. This figure provides the comparison of two proposed algorithms, where AOBWS refers to alternating optimization for backscatter aided wireless powered sensors, which provide optimal EE by optimizing both CE transmit power and reflection coefficient of roadside sensor through the proposed alternating optimization framework. AOBWS optimizes CE transmit power under given reflection coefficient of roadside sensor in the first stage while in the 2nd stage it optimizes reflection coefficient of roadside sensor under optimal CE transmit power obtained in the 1st stage. While OCETP refers to the first stage of AOBWS which obtains optimal CE transmit power under given reflection coefficient of roadside sensor. From the figure, it can be noticed that our proposed AOBWS has higher EE after performing its 2 stage operation as compare to OCETP. 

\begin{figure}
	\centering
	\includegraphics[width=85mm]{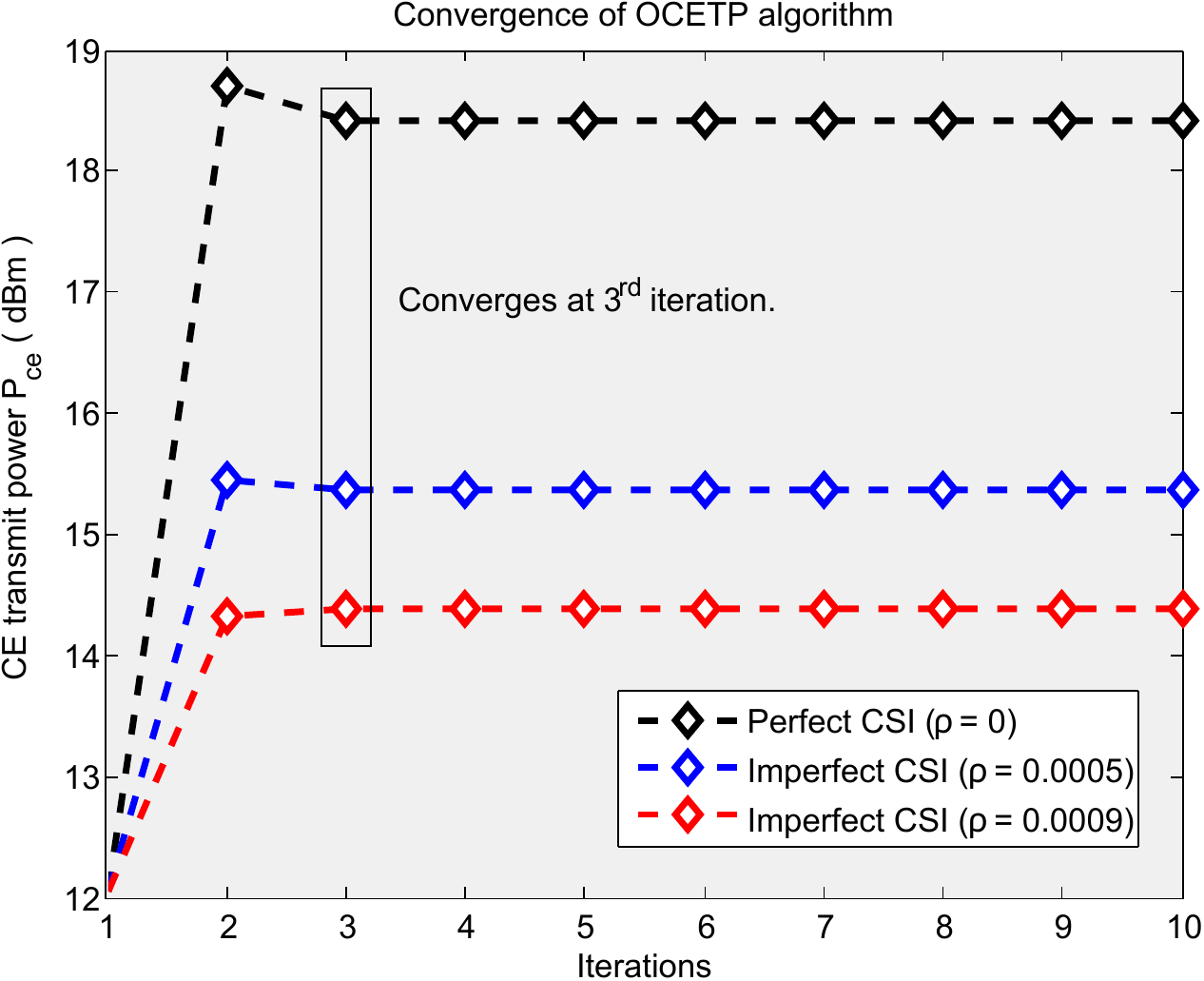}
	\caption{Convergence of OCETP algorithm under various relative channel errors $\rho$.}
	\label{FIG:4}
\end{figure}
\begin{figure}
	\centering
	\includegraphics[width=80mm]{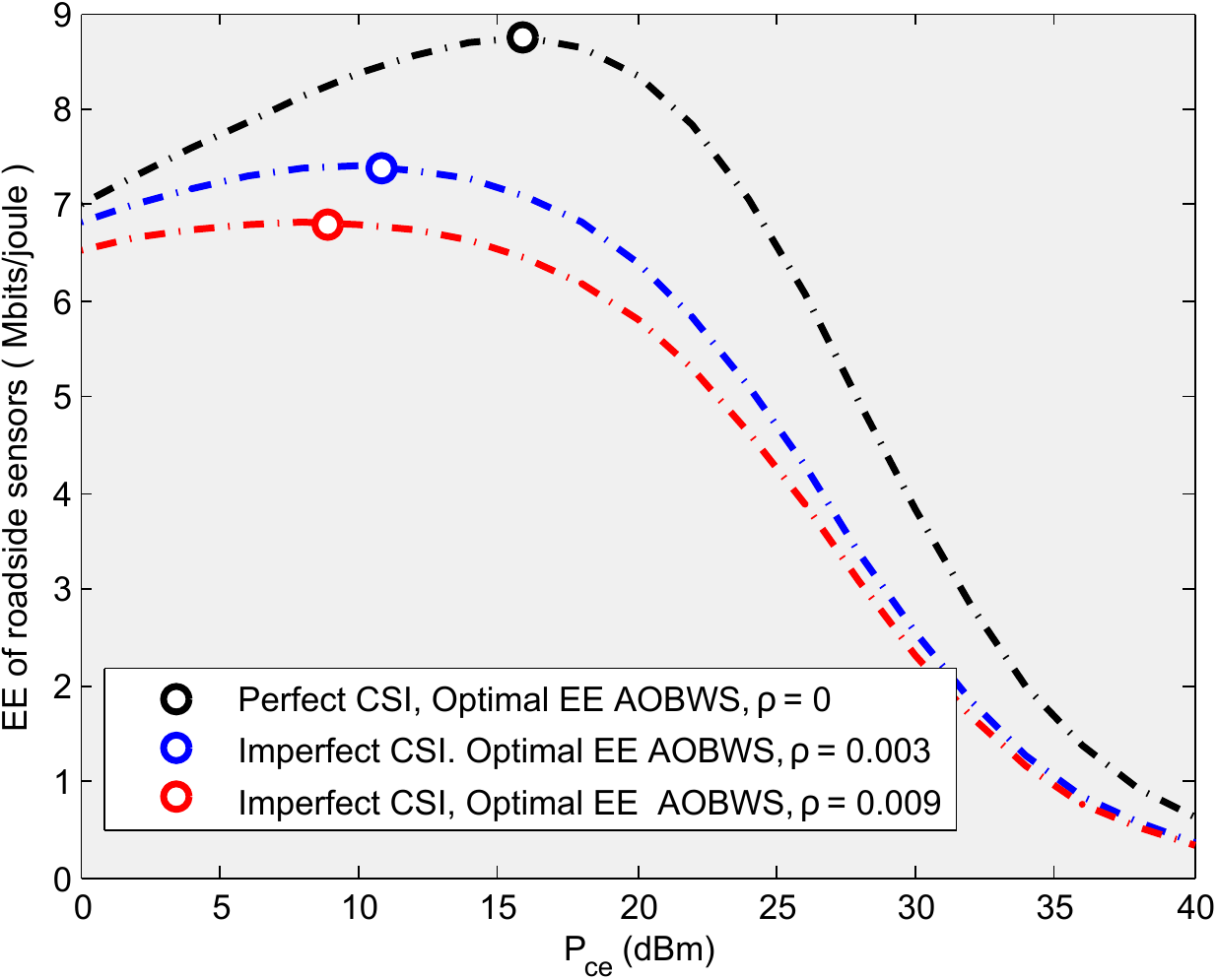}
	\caption{Optimal EE through proposed AOBWS algorithm under various relative channel errors $\rho$.}
	\label{FIG:5}
\end{figure}

As our proposed alternating optimization framework (AOBWS) consists of two stages, where the first stage employs an iterative algorithm for optimal CE transmit power named OCETP while in 2nd stage it uses a non-iterative algorithm for optimal reflection coefficient of RS. Therefore, it is vital to analyze the convergence of the OCETP algorithm. In Fig.~\ref{FIG:4}, EE convergence of OCETP algorithm versus iterations is demonstrated with different relative channel errors ($\rho$). The obtained result presents that the OCETP algorithm usually converges in three iterations regardless of $\rho$. It is observed that $\rho$ affects EE, but their influence on the convergence of the OCETP algorithm is almost negligible.

Fig.~\ref{FIG:5} depicts optimal EE obtained by the proposed AOBWS algorithm under different relative channel errors. It can be observed from the figure that EE as a function of $P_{ce}$  first increases and then decreases with increasing $P_{ce}$. It is because when $P_{ce}$ is increased beyond the optimal $P^*_{ce}$, the RS sum-rate increases very slowly relative to the power consumption. It can also be examined from the figure that higher relative channel errors result in higher EE degradation.

\begin{figure}
	\centering
	\includegraphics[width=95mm]{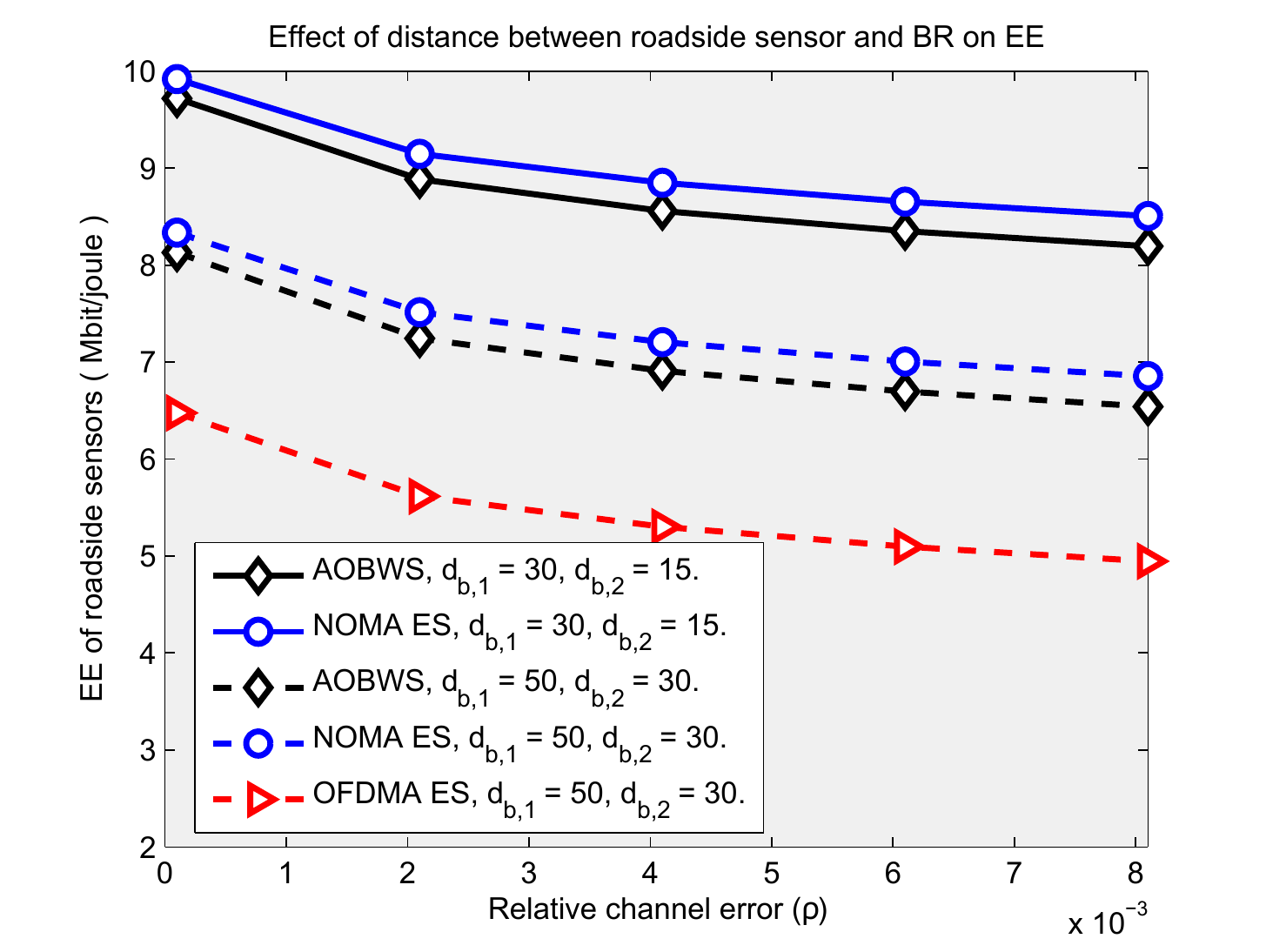}
	\caption{EE of roadside sensors versus $\rho$ under different distances between roadside sensor and BR. }
	\label{FIG:6}
\end{figure}

Fig.~\ref{FIG:6} compares the EE of the proposed AOBWS algorithm with the global optimal ES algorithm and OFDMA. The comparison is done versus relative channel errors while considering different distances between RS and BSR installed on RSU. It can be analyzed from the figure that AOBWS obtains very close EE performance to the ES algorithm with very low complexity. Moreover, it can be observed that higher relative channel error has a higher degrading effect on the EE. From the figure, it can also be noticed that a greater distance between RSs and BSR installed on RSU  results in higher EE degradation.

\begin{figure}
	\centering
	\includegraphics[width=95mm]{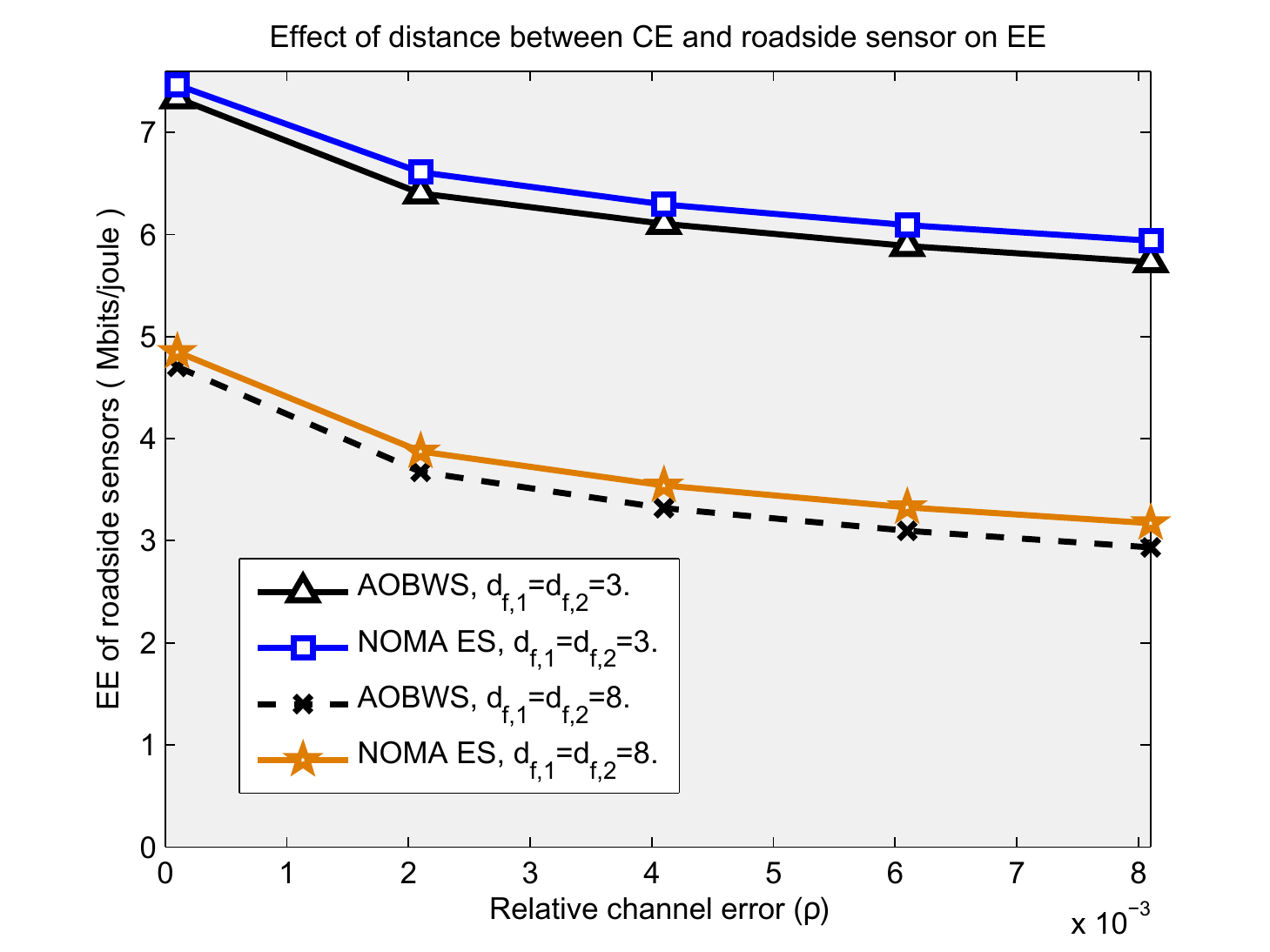}
	\caption{EE of roadside sensors versus $\rho$ under different distances between CE and roadside sensor. }
	\label{FIG:7}
\end{figure}

In Fig.~\ref{FIG:7}, the EE of the proposed AOBWS algorithm is compared with the global optimal ES algorithm versus relative channel errors. This figure presents the effect of distance between dedicated CE and RS. From the results, It can be examined that a higher distance between CE and RS will result in lower EE. Moreover, the proposed AOBWS algorithm shows very close performance with the ES algorithm, which presents the efficacy of our proposed algorithm with low computational complexity.

\begin{figure} 
	\centering
	\includegraphics[width=85mm]{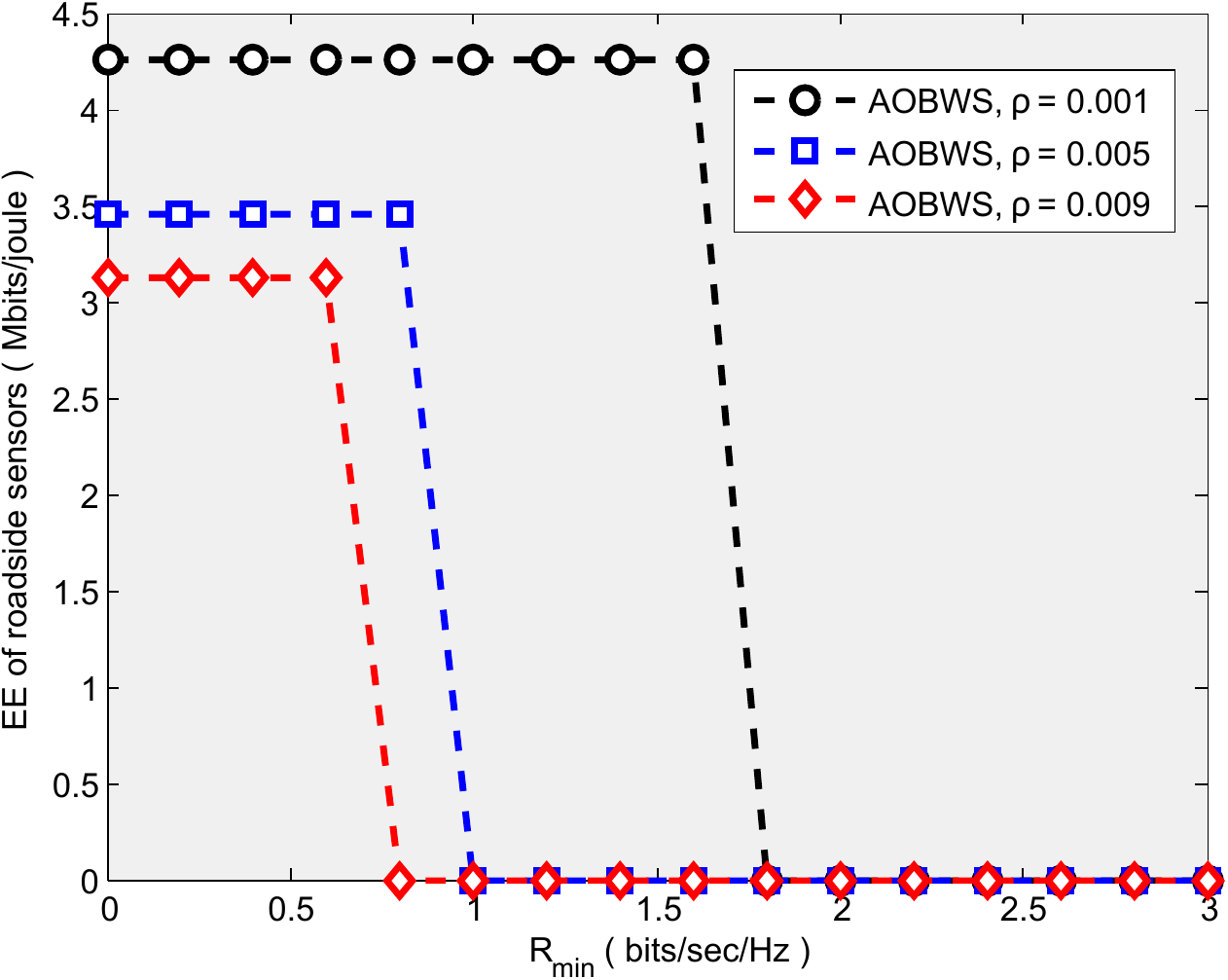}
	\caption{EE of roadside sensors versus $R_{min}$  }
	\label{FIG:8}
\end{figure}

Fig.~\ref{FIG:8} shows the EE of RS versus QoS requirement of RS in term minimum data rate ($R_{min}$) for proposed AOBWS algorithm. The analysis of this figure is done with $d_{f,1}=10$, $d_{f,2}=5$, $d_{b,1}=50$, and $d_{b,2}=30$. This figure depicts that  higher relative channel error will result in lower guaranteed data rate of the network. 


\section{Conclusion}
In this article, we proposed an alternating optimization framework for backscatter aided wireless powered roadside sensors to infrastructure communications with imperfect CSI. By considering various QoS requirements of bistatic backscatter communications for roadside sensors, we aim to maximize the EE of the network. The EE is maximized under channel uncertainties by optimizing transmit power of CE and reflection coefficient of RS, with very low computational complexity for practical implementations. The proposed problem for the solution is decoupled into two stages, which yields the proposed AOBWS algorithm. The complexity analysis of the proposed algorithm is analyzed in detail and is compared with the global optimal ES algorithm. It is observed that our proposed algorithm can obtain near-optimal EE performance with very low complexity. Moreover, numerical results present the efficacy of our proposed result in terms of computational complexity and EE performance. In the future, we want to extend our work to a multi-cluster of RSs , full-duplex (FD) RSs, and multi-antenna RSs.

\appendices


\section{Concavity of $\overline{R}$ with respect to $P_{ce}$}
Appendix A presents the concavity of $\overline{R}$ with respect to $P_{ce}$. The first derivative of $\overline{R}$ w.r.t. $P_{ce}$ is given as
\begin{align}
& \frac {\partial \overline{R}}{\partial P_{ce}} = \frac {BWT_{t,1}\Pi_{1}\sigma^2_n}{\ln(2)P_{ce}X}+\frac{BWT_{t,2}\Pi_{2}\sigma^2_n}{\ln(2)P_{ce}Y}. \label{eq 36}
\end{align}
The second order derivative is given as
\begin{align}
& \frac {\partial^2 \overline{R}}{\partial {P_{ce}}^2} =-\frac {V}{\ln(2){P_{ce}}^2X^2
}-\frac{W}{\ln(2){P_{ce}}^2Y^2}. \label{eq 37}
\end{align}
where $V=BWT_{t,1}\Pi_{1}({\sigma^2_n})({2\sum_{k=1}^{2}\sigma^{2}_{e}P_{ce}\Gamma_k+\sigma^2_n})$ and $W=BWT_{t,2}\Pi_{2}({\sigma^2_n})({2({P_{ce}\Gamma_1|\hat{H}_{1}|^2}+\sum_{k=1}^{2}\sigma^{2}_{e}P_{ce}\Gamma_k)+\sigma^2_n})$. As second order derivative of $\overline{R}$ i.e. $\frac {\partial^2 \overline{R}}{\partial {P_{ce}}^2} < 0 $, therefore
$\overline{R}$ is concave and is an increasing function of $P_{ce}$.

\section{Derivation of closed-form solution of optimal $P_{ce}$}

We exploit the KKT conditions such as

\begin{align}
& \frac {\partial \mathit{L(P_{ce},\boldsymbol{\mu},\lambda,\boldsymbol{\beta})}}{\partial P_{ce}}|_{P_{ce}=P^*_{ce}} = 0. \label{eq 38}
\end{align}
The above equation results in
\begin{align}
\frac{A}{\ln(2)P_{ce}\big(BP_{ce}+\sigma^2_n\big)}+\frac{C}{\ln(2)P_{ce}\big(DP_{ce}+\sigma^2_n\big)}+G=0,  \label{eq 39}
\end{align}
where
\begin{align}
	& A = T_{t,1}BW\Pi_{1}\sigma^2_n, \label{eq 40} \\
	& B = \sum_{k=1}^{2}\sigma^{2}_{e}\Gamma_k, \label{eq 41} \\
	& C = T_{t,2}BW\Pi_{2}\sigma^2_n, \label{eq 42} \\
	& D = \Gamma_1|\hat{H}_{1}|^2+\sum_{k=1}^{2}\sigma^{2}_{e}\Gamma_,k \label{eq 43}\\
	& G = \sum_{k=1}^{2}\mu_k Q_k+\sum_{k=1}^{2}\beta_kP^H_k- \sum_{k=1}^{2}\frac{\psi}{\varkappa_k}-\lambda, \label{eq:44}
\end{align}
\begin{equation}
 Q_k =\Gamma_k|\hat{H}_{k}|^2-\bigg(2^{\frac{R_{min}-T_{t,k}\Phi_k}{T_{t,k}\Pi_k}}\bigg)\bigg({\sum_{l=1}^{k-1}\Gamma_l|\hat{H}_{l}|^2}+{\sum_{k=1}^{2}\sigma^{2}_{e}\Gamma_k}\bigg), \label{eq:45}
\end{equation}
and,
\begin{equation}
P^H_k = \xi(1-\Gamma_k)|\hat{H}_{f,k}|^2T_{t,k}+\xi\Gamma_kT_{h,k}. \label{eq: 46}
\end{equation}
After some computations
\begin{align}
&(\ln(2)BDG)P_{ce}^3+(\ln(2)BG\sigma^2_n+\ln(2)DG\sigma^2_n)P_{ce}^2+\nonumber\\
&(\ln(2)G(\sigma^2_n)^2+AD+CB)P_{ce}+(A\sigma^2_n+C\sigma^2_n)=0. \label{eq 46}
\end{align} 
The above equation can be solved through Cardano's formulae as follow 

\begin{align}
P^*_{ce}= &\sqrt[3]{q+\sqrt{q^2+\big(r-p^2\big)^3}}+\sqrt[3]{q-\sqrt{q^2+\big(r-p^2\big)^3}}+p, 
\end{align}
where, 
\begin{align}
& p= \frac{-b}{3a},\\
& q = p^3+ \frac{bc-3ad}{6a^2}, \\
& r = \frac{c}{3a},
\end{align}
and
\begin{align}
& a= \ln(2)BDG,\\
& b = \ln(2)BG\sigma^2_n+\ln(2)DG\sigma^2_n,\\
& c = \ln(2)G(\sigma^2_n)^2+AD+CB ,\\
& d = A\sigma^2_n+C\sigma^2_n.
\end{align}

\section{Concavity of $\overline{R}$ with respect to reflection coefficients of RSs}
In the optimization problem defined in Eq. (\ref{eq:32}) , we will present that $\overline{R}$ is concave and is increasing function of reflection coefficient of RS. A function is concave, if its Hessian matrix is negative definite. The hessian matrix is negative definite, if its all eigenvalues are negative. Here we derive a Hessian matrix
of $\overline{R}$ and demonstrate it as negative
definite. The $\overline{R}$ in optimization problem in Eq. (\ref{eq:32})  can be written as  

\begin{align}
\overline{R}= &T_{t,1}BW(\Pi_{1}\log_{2}\bigg(\frac{P_{ce}\Gamma_1|\hat{H}_{1}|^2}{\sigma^{2}_{e} P_{ce}\theta+\sigma^2_n}\bigg)+\Phi_{1})+\\
&T_{t,2}BW(\Pi_{2}\log_{2}\bigg(\frac{P_{ce}\Gamma_2|\hat{H}_{2}|^2}{{P_{ce}\Gamma_1|\hat{H}_{1}|^2}+\sigma^{2}_{e} P_{ce}\theta+\sigma^2_n}\bigg)+\Phi_{2}). \nonumber\label{eq 100}
\end{align} 
The Hessian matrix of above function with respect to $\Gamma_{1}$ and  $\Gamma_{2}$ is defined as

\begin{align}
H=
\begin{bmatrix}
\frac {\partial \overline{R}}{\partial^2 \Gamma_{1}} & \frac {\partial \overline{R}}{\partial \Gamma_{1}\partial \Gamma_{2}} \\
\frac {\partial \overline{R}}{\partial \Gamma_{2}\partial \Gamma_{1}} & \frac {\partial \overline{R}}{\partial^2 \Gamma_{2}} 
\end{bmatrix},
\end{align}
where
\begin{align}
\frac {\partial \overline{R}}{\partial^2 \Gamma_{1}} & = \Upsilon_{1,1} \nonumber\\ 
&= -\frac{T_{t,1}BW\Pi_{1}}{\ln (2) \Gamma_{1}^2}+ \frac{T_{t,2}BW\Pi_{2}(P_{ce}|\hat{H}_{1}|^2)^2}{\ln(2)\Big(P_{ce}\Gamma_1|\hat{H}_{1}|^2+(\sigma^{2}_{e} P_{ce}\theta+\sigma^2_n)^2\Big)},
\end{align}

\begin{align}
\frac {\partial \overline{R}}{\partial \Gamma_{1}\partial \Gamma_{2}} = \Upsilon_{1,2} = 0,
\end{align}

\begin{align}
\frac {\partial \overline{R}}{\partial \Gamma_{2}\partial \Gamma_{1}} = \Upsilon_{2,1} = 0,
\end{align}
and
\begin{align}
\frac {\partial \overline{R}}{\partial^2 \Gamma_{2}} = \Upsilon_{2,2} = -\frac{T_{t,2}BW\Pi_{2}}{\ln(2)\Gamma_{2}^2}.
\end{align}
Now the eigenvalues of Hessian matrix can be calcualted as follow 
\begin{align}
\begin{bmatrix}
\Upsilon_{1,1} & \Upsilon_{1,2} \\
\Upsilon_{2,1} & \Upsilon_{2,2}
\end{bmatrix}
-
\begin{bmatrix}
\pi & 0 \\
0 & \pi
\end{bmatrix}.
\end{align}
\begin{align}
\det
\begin{pmatrix}
\Upsilon_{1,1}-\pi & \Upsilon_{1,2} \\
\Upsilon_{2,1} & \Upsilon_{2,2}-\pi
\end{pmatrix},
\end{align}
det of above matrix results in,
\begin{align}
\Upsilon_{1,1}\Upsilon_{2,2}-\pi\Upsilon_{1,1}-\pi\Upsilon_{2,2}+\pi^2-\Upsilon_{1,2}\Upsilon_{2,1} = 0.
\end{align}
After writing it in standard form of $Mx^2 + Nx + O$, 
\begin{align}
\pi^2- (\Upsilon_{1,1}+\Upsilon_{2,2})\pi + (\Upsilon_{1,1}\Upsilon_{2,2}-\Upsilon_{1,2}\Upsilon_{2,1})  = 0,
\end{align}
where $M$=1, $N$=$\Upsilon_{1,1}+\Upsilon_{2,2}$, $O$=$\Upsilon_{1,1}\Upsilon_{2,2}-\Upsilon_{1,2}\Upsilon_{2,1}$
Then, 
The solution of above problem is as follow
\begin{align}
\pi = \frac{-N \pm \sqrt{N^2-4MO}}{2M}.
\end{align}	
As, the two possible eigenvalues computed through above formulae is always negative for the hessian matrix, therefore $\overline{R}$ is concave.


 \vskip -2\baselineskip
\begin{IEEEbiography}[{\includegraphics[width=1in,height=1.25in,clip,keepaspectratio]{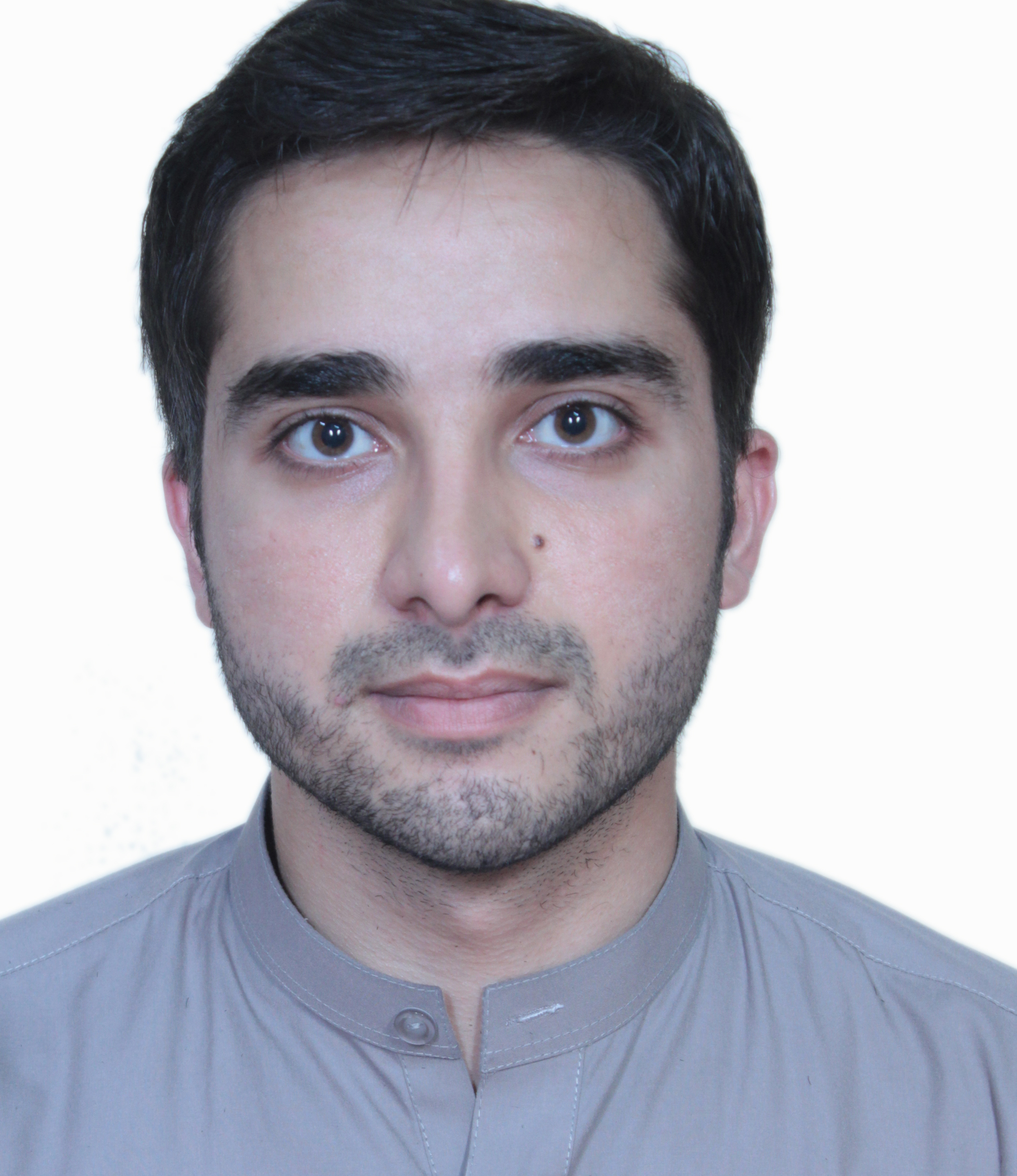}}]{Asim Ihsan}
	received the B.S. degree in Telecommunication Engineering from the University of Engineering and Technology (UET), Peshawar, Pakistan, in 2015, and the M.S. degree in Information and Communication Engineering from Xi’an Jiaotong University (XJTU), Xi’an, China, in 2018. He is
	currently pursuing the Ph.D. degree in Information and Communication Engineering with Shanghai Jiao Tong University (SJTU), Shanghai, China. His current research interests include the internet of vehicles, backscatter communications, physical layer security, and wireless sensor networks. He is an active reviewer of peer reviewed international journals, such as IEEE, Elsevier, Springer, and Wiley.
\end{IEEEbiography}
\vskip -2\baselineskip plus -1fil 
\begin{IEEEbiography}[{\includegraphics[width=1in,height=1.25in,clip,keepaspectratio]{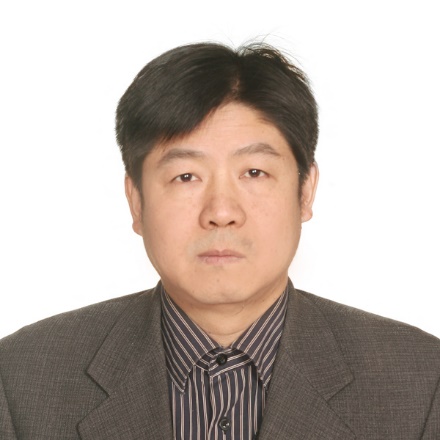}}]{Wen Chen}
	(Senior Member, IEEE) is a tenured Professor with the Department of Electronic Engineering, Shanghai Jiao Tong University, China, where he is the director of Broadband Access Network Laboratory. He is a fellow of Chinese Institute of Electronics and the distinguished lecturers of IEEE Communications Society and IEEE Vehicular Technology Society. He is the Shanghai Chapter Chair of IEEE Vehicular Technology Society, an Editors of IEEE Transactions on Wireless Communications, IEEE Transactions on Communications, IEEE Access and IEEE Open Journal of Vehicular Technology. His research interests include multiple access, wireless AI and meta-surface communications. He has published more than 100 papers in IEEE journals and more than 100 papers in IEEE Conferences, with citations more than 6000 in google scholar.
\end{IEEEbiography}
\vskip -2\baselineskip plus -1fil
\begin{IEEEbiography}
	[{\includegraphics[width=1in,height=1.5in,clip,keepaspectratio]{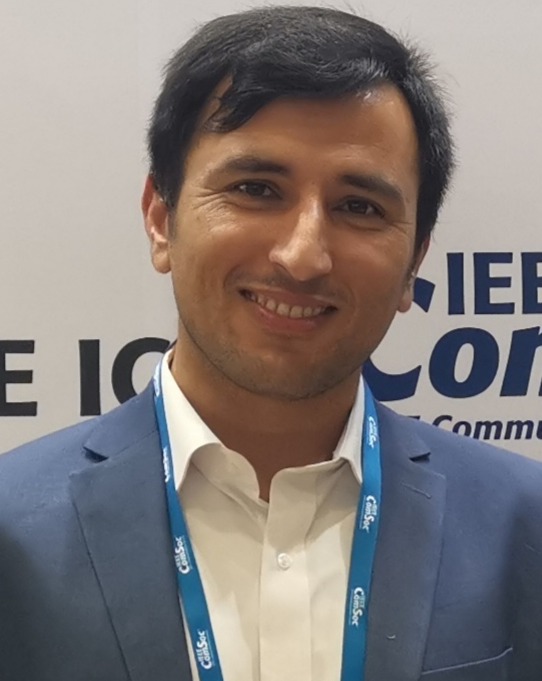}}]{Wali Ullah Khan} 
	received the Master degree in Electrical Engineering from COMSATS University Islamabad, Pakistan, in 2017, and the Ph.D. degree in Information and Communication Engineering from Shandong University, Qingdao, China, in 2020. He is currently working with the Interdisciplinary Centre for Security, Reliability and Trust (SnT), University of Luxembourg, Luxembourg. He has authored/coauthored more than 50 publications, including international journals, peer-reviewed conferences, and book chapters. His research interests include convex/nonconvex optimizations, non-orthogonal multiple access, reflecting intelligent surfaces, ambient backscatter communications, Internet of things, intelligent transportation systems, satellite communications, physical layer security, and applications of machine learning.
\end{IEEEbiography}
\vskip -2\baselineskip plus -1fil
\begin{IEEEbiography}
	[{\includegraphics[width=1in,height=1.5in,clip,keepaspectratio]{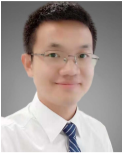}}]{Qingqing Wu} (Senior Member, IEEE) received the B.Eng. degree in electronic engineering from the South China University of Technology in 2012 and the Ph.D. degree in electronic engineering from Shanghai Jiao Tong University (SJTU) in 2016. He is currently an assistant professor with the State key laboratory of Internet of Things for Smart City, University of Macau. From 2016 to 2020, he was a Research Fellow in the Department of Electrical and Computer Engineering at National University of Singapore. His current research interest includes intelligent reflecting surface (IRS), unmanned aerial vehicle (UAV) communications, and MIMO transceiver design. He has coauthored more than 150 IEEE papers with 24 ESI highly cited papers and 8 ESI hot papers, which have received more than 10,000 Google citations. He was listed as Clarivate ESI Highly Cited Researcher in 2021 and World’s Top 2\% Scientist by Stanford University in 2020.
	
	He was the recipient of the IEEE Communications Society Young Author Best Paper Award in 2021, the Outstanding Ph.D. Thesis Award of China Institute of Communications in 2017, the Outstanding Ph.D. Thesis Funding in SJTU in 2016, the IEEE ICCC Best Paper Award in 2021, and IEEE WCSP Best Paper Award in 2015. He was the Exemplary Editor of IEEE Communications Letters in 2019 and the Exemplary Reviewer of several IEEE journals. He serves as an Associate Editor for IEEE Transactions on Communications, IEEE Communications Letters, IEEE Wireless Communications Letters, IEEE Open Journal of Communications Society (OJ-COMS), and IEEE Open Journal of Vehicular Technology (OJVT). He is the Lead  Guest Editor for IEEE Journal on Selected Areas in Communications on ” UAV Communications in 5G and Beyond Networks”, and the  Guest Editor for IEEE  OJVT on  “6G  Intelligent Communications” and IEEE OJ-COMS on “Reconfigurable  Intelligent Surface Based Communications for 6G  Wireless  Networks”. He is the workshop co-chair for IEEE ICC 2019-2022 workshop on “Integrating UAVs into 5G and Beyond”, and the workshop co-chair for IEEE GLOBECOM 2020 and ICC 2021 workshop on “Reconfigurable Intelligent Surfaces for Wireless Communication for Beyond 5G”. He serves as the Workshops and Symposia Officer of Reconfigurable Intelligent Surfaces Emerging Technology Initiative and Research Blog Officer of Aerial Communications Emerging Technology Initiative. He is the IEEE Communications Society Young Professional Chair in Asia Pacific Region.	
\end{IEEEbiography}
\vskip -2\baselineskip plus -1fil
\begin{IEEEbiography}
	[{\includegraphics[width=1in,height=1.5in,clip,keepaspectratio]{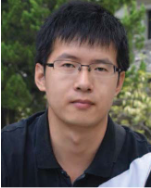}}]{Kunlun Wang} 
	(Member, IEEE) received the Ph.D. degree in electronic  engineering from Shanghai Jiao Tong University, Shanghai, China, in 2016. From 2016 to 2017, he was with Huawei Technologies Company Ltd., where he was involved in energy efficiency algorithm design. From 2017 to 2019, he was with the Key Laboratory of Wireless Sensor Network and Communication, SIMIT, Chinese Academy of Sciences, Shanghai. From 2019 to 2020, he was with the School of Information Science and Technology, Shanghai Tech University. Since 2021, he has been a Professor with the School of Communication and Electronic Engineering, East China Normal University. His current research interests include energy efficient communications, fog computing networks, resource allocation, and optimization algorithm.  
\end{IEEEbiography}

\end{document}